%% file: main.tex
\documentclass[fleqn,usenatbib]{mnras}

\usepackage{newtxtext,newtxmath}

\usepackage[T1]{fontenc}
\DeclareRobustCommand{\VAN}[3]{#2}
\let\VANthebibliography\thebibliography
\def\thebibliography{\DeclareRobustCommand{\VAN}[3]{##3}\VANthebibliography}

\usepackage{subfiles}
\usepackage{float}
\usepackage{amsmath}
\usepackage{booktabs}
\usepackage{caption}
\usepackage{subcaption}
\usepackage{physics}
\usepackage{array}
\usepackage{cuted}
\usepackage{multirow}
\usepackage{stfloats}
\usepackage{widetext}
\usepackage{hyperref}
\usepackage{ae,aecompl}

\usepackage{graphicx}	
\usepackage{amsmath}	

\usepackage{acronym}
\newacro{GW}[GW]{gravitational wave}
\newacro{CBC}[CBC]{compact binary coalescence}
\newacro{LVK}[LVK]{LIGO-Virgo-KAGRA}
\newacro{LVKI}[LVKI]{LIGO-Virgo-KAGRA-LIGO India}
\newacro{ET}[ET]{Einstein Telescope}
\newacro{CE}[CE]{Cosmic Explorer}
\newacro{LVKI+ET}[LVKI+ET]{LIGO-Virgo-KAGRA-LIGO India + Einstein Telescope}
\newacro{LVKI+ET+CE}[LVKI+ET]{LIGO-Virgo-KAGRA-LIGO India + Einstein Telescope + Cosmic explorer}

\newacro{SGWB}[SGWB]{stochastic gravitational-wave background} 

\newacro{CMB}[CMB]{cosmic microwave background}

\def \be {\begin{equation}}
\def \ee {\end{equation}}
\def \GW {{\rm {GW}}}
\def \ve {{\bf \hat{e}}}

\def \vtheta {\boldsymbol{\theta}}
\def \ra {\texttt{ra} }
\def \dec {\texttt{dec} }

\title[GWB--galaxies cross-correlation forecasts]{Prospects for multi-messenger discovery of the gravitational-wave background anisotropies via cross-correlation with galaxies}

\author[R. Bertrand-Delgado et al.]{
Raphael Bertrand-Delgado\thanks{E-mail: raphael.bertrand@uzh.ch },
Felipe Andrade-Oliveira,
Michael Ebersold,
Marcelle Soares-Santos
\\
Physik-Institut, University of Zurich, Winterthurerstrasse 190, 8057 Zurich, Switzerland
}

\date{Accepted XXX. Received YYY; in original form ZZZ}

\pubyear{\the\year{}}

\begin{document}
\label{firstpage}
\pagerange{\pageref{firstpage}--\pageref{lastpage}}
\maketitle

\begin{abstract}
We present new empirically grounded forecasts for the observability of  
stochastic gravitational-wave background anisotropies assuming a population of stellar-mass compact binary coalescences as its source. We quantified the discovery potential using simulations based on the Euclid Flagship Galaxy Catalogue and LIGO-Virgo-KAGRA observational constraints in combination with detailed theoretical modelling. We considered the multi-messenger cross-correlation with galaxies as well as the gravitational wave-only cross-correlation across observation-time bins. 
We found that signal-to-noise ratio greater than five is achievable via cross-correlation of a galaxy catalogue that is complete up to limiting magnitude $i < 24.7$ and has redshift uncertainties $\sigma_z = 0.003 (1+z)$ with a gravitational-wave sky map corresponding to 2 years of observation  
by a third-generation detector network comprising three  interferometers. Network configurations with LIGO-Virgo-KAGRA plus up to two third-generation detectors would be limited to moderate signal-to-noise levels, between three and five, except if merger rates are near the upper bound set by observational constraints, in which case the threshold of five would be reached after 4 years. Detectability without the multi-messenger tracer has proven significantly more challenging, with signal-to-noise ratio remaining below three after 10 years.
In light of the plans being carried out in the community for ongoing and upcoming galaxy surveys, this work bodes well for the multi-messenger discovery and exploration of the stochastic gravitational-wave background in the era of next-generation observatories such as the Einstein Telescope and Cosmic Explorer.
\end{abstract}




\section{Introduction}
\subfile{introduction}

\section{Simulations}
\subfile{simulations}

\section{Estimators}
\subfile{estimators}

\section{Modelling the angular power spectrum}
\subfile{prediction}

\section{Results}
\subfile{results}

\section{Conclusions}

\subfile{conclusion}

\section*{Acknowledgements}

We are grateful to the anonymous referee for comments that have led to a much improved manuscript. We thank Haowen Zhong for useful comments on the paper draft.
Funding for this work is provided by the University of Zurich (UZH). ME and RBD received support from the UZH Postdoc Grant No. [FK-25-109] and the UZH Candoc Grant No. [FK-26-081], respectively.

\section*{Data availability}
The simulated \ac{CBC} populations are available upon reasonable request. The Euclid Flagship 2 simulated galaxy catalogue is publicly available at: \href{https://cosmohub.pic.es/catalogs/353}{https://cosmohub.pic.es/catalogs/353}. 
The \ac{CE} location is taken from \href{https://cosmicexplorer.org/celocations.html}{https://cosmicexplorer.org/celocations.html}. The PSDs for the \ac{LVK}, \ac{ET}, and \ac{CE} are respectively available at \href{https://dcc.ligo.org/LIGO-T2200043/public}{https://dcc.ligo.org/LIGO-T2200043/public}, \href{https://apps.et-gw.eu/tds/?r=18213}{https://apps.et-gw.eu/tds/?r=18213}, and \href{https://cosmicexplorer.org/sensitivity.html}{https://cosmicexplorer.org/sensitivity.html}.



\bibliographystyle{mnras}
\bibliography{ref}


\appendix

\subfile{appendix}


\bsp	
\label{lastpage}
\end{document}

%% file: introduction.tex
In the decade since the first direct observation of \acp{GW} from a \ac{CBC} \citep{abbott_observation_2016}, hundreds of such events have been catalogued by the \ac{LVK} network of \ac{GW} observatories \citep{gwtc5_observation}. These detections imply the existence of a \ac{SGWB} made of the superposition of dim and unresolved sources accumulated over the history of the universe \citep{christensen_stochastic_2018, renzini_stochastic_2022}. 
In addition to astrophysical sources such as \acp{CBC}  \citep{zhu_stochastic_2011,regimbau_astrophysical_2011,meacher_statistical_2014}, other expected \ac{SGWB} sources include inflation and first order phase transitions in the early universe \citep{mukherjee_scalar-induced_2026,gammal_reconstructing_2025, campeti_measuring_2021, dimastrogiovanni_primordial_2017, sakellariadou_cosmic_2009,jenkins_anisotropies_2018-1, hindmarsh_phase_2021,caprini_cosmological_2018, ricciardone_cross-correlating_2021}. 
Astrophysical sources are expected to feature anisotropies that follow the distribution of large-scale structures 
\citep{cusin_anisotropy_2017, jenkins_anisotropies_2018, jenkins_anisotropies_2019, contaldi_anisotropies_2017}.
Therefore, the angular power spectrum of the \ac{SGWB} is a promising observable to study the evolution of compact binaries and the large-scale structures they inhabit \citep{cusin_first_2018, jenkins_estimating_2019, pedrotti_cosmology_2025}. 

Searches for the \ac{SGWB} have been vigorously pursued by the \ac{LVK}, with sensitivity steadily approaching the detection threshold \citep{gwtc5_isotropic_limit}. Meanwhile, next-generation \ac{GW} observatories such as the \ac{ET} \citep{maggiore_science_2020} and \ac{CE}  \citep{evans_horizon_2021} are being planned to make a leap in sensitivity and angular resolution relative to the \ac{LVK}, enabling confident detection of the \ac{SGWB} signal and its anisotropies. 

Predictions for the \ac{SGWB} angular power spectrum previously reported in the literature include theoretical modelling of  sources and noise \citep{cusin_properties_2019, jenkins_estimating_2019, alonso_detecting_2020, alonso_tomographic_2024, bellomo_class_gwb_2022, capurri_intensity_2021, capurri_detectability_2022}.  Investigations using simulated realisations of discrete source populations  can facilitate the assessment of observational effects that pose a challenge to analytical modelling in preparation for future analyses of observed data \citep{yang_measurement_2023, yang_searching_2020}. The success of this approach, however, requires simulations capable of accurately populating a large sample of galaxies covering a wide area of the sky with \ac{GW} sources following their evolution up to high redshifts. A first exploration 
of this galaxy-informed approach 
has recently been reported \citep{yang_textiteuclid_2025}.

Motivated by this scenario,
we present a new study that combines analytical modelling and simulations to estimate, under realistic assumptions, the \ac{SGWB} anisotropies in the \ac{LVK} frequency band assuming a population of \acp{CBC} as its source. 
We develop a framework to generate and distribute \acp{CBC} onto an input host galaxy catalogue and compute the corresponding \ac{SGWB} sky map within a given observing time window. We then analyse the sky-maps to extract the signal and its uncertainties from the \ac{SGWB} cross-correlation with an observed galaxy distribution as well as from the cross-correlation between observation time window bins. 
The analysis is first performed in the shot-noise limited regime and then considering the instrumental noise expected for several possible detector network scenarios,  using the  \ac{LVK}, \ac{ET}, and \ac{CE} target sensitivity curves. Our shot-noise derivations, while analogous to \cite{bellomo_class_gwb_2022}, are adapted to a galaxy-hosted CBC population instead of larger haloes. This facilitates comparison with simulations.
We also derive physically motivated analytical expressions for the angular power spectra and their covariance.
We assume \ac{CBC} mass distributions and merger rates reported by the LVK based on observations up to the first part of the fourth observing run (O4a) \citep{abac_gwtc-40_2025, abac_upper_2025}, and we use galaxies from the Euclid Flagship Simulation Catalogue \citep{castander_euclid_2025}. 

In their study, \cite{yang_textiteuclid_2025} also used the Flagship catalogue and LVK constraints, albeit from the third observing run (O3) \citep{abbott_gwtc-3_2023}. Beyond using more recent publicly available (O4a) constraints, our work further advances the field by presenting a more empirically grounded approach. Notably, we model evolution of the merger rates accounting for the growing uncertainty with redshift, and we consider a finite observing time window over which only a certain amount of events would occur (to obtain a more realistic and accurately modelled shot noise). We also build a full sky map that ensures that the $\ell$-modes are uncorrelated resulting in a diagonal covariance matrix in the spherical harmonic basis. Moreover, for the cross-correlation analysis, specifically, we create a mock observed galaxy catalogue based on expected observational capabilities of the next generation galaxy surveys. We also cross-check our derived covariance with simulations and analytical calculations. Our analytical model derived for the kernel is physically motivated, with its redshift dependence derived from the estimated merger rate and the energy flux evolution.
Finally, we also consider the instrumental noise.

This paper is structured as follows: Section~\ref{sec:simulations} describes the simulations, Section~\ref{sec:estimators} covers the estimators used to obtain an unbiased cross-correlation and compute the shot noise, Section~\ref{sec:modelling} describes the modelling used in the analysis, and Section \ref{sec:results} presents our results.  
We adopt the fiducial Flagship cosmology
namely, a flat $\Lambda$CDM cosmology with $\Omega_m=0.319$, $\Omega_b=0.049$, $\Omega_\Lambda + \Omega_\gamma=0.681$, $A_s = 2.1\times 10^{-9}$, $n_s=0.96$, $h = 0.67$.

%% file: simulations.tex
\label{sec:simulations}
\label{sec:simulations_sky_maps} 

\subsection{Preparation of the galaxy catalogues}

We use the publicly available Euclid Flagship Simulation Catalogue 
\citep{castander_euclid_2025}, a set of synthetic galaxy light cones covering approximately one sky octant and spanning the range $0 <z < 3$.
The Flagship catalogue was  constructed from large-volume N-body simulations where galaxies were added to halos through halo abundance matching \citep{lehmann_concentration_2017}.
Modelled galaxy properties include luminosity and flux in several bands, cosmological and observed redshifts going up to $z =3$, sky positions, stellar masses from $10^5$ to $10^{12} M_\odot /h^2$, velocities, spectral energy distributions, shapes and sizes, star formation rates, metallicities, emission line fluxes, and lensing properties.
With the exception of a small patch of $5\times5$ ${\rm deg}^2$ where fainter objects are included, Flagship is mostly uniform in depth, with completeness up to apparent magnitude $H<26.6$. 

As GW sources can be located anywhere in the sky,  
we produce an all-sky catalogue from the original Flagship octant. For each galaxy, we perform a random number of reflections between 0 and 7 on its \texttt{ra} and \texttt{dec} coordinates. By doing so, we randomly distribute each galaxy in the eight octants of the sky, but conserve their clustering. Spurious clustering caused by this manoeuvrer 
does not affect our results, as they arise at very large scales which are conservatively excluded from our analysis ($\ell < 8$).

In this paper, the Flagship catalogue is used in two ways. First, we use it as the basis for our {\it host galaxy catalogue}, to ensure that we produce \ac{SGWB} maps with the correct clustering properties. Second, we use it to create a mock {\it observed galaxy catalogue}, for the cross-correlation analysis. 

We build our host galaxy catalogue using all Flagship galaxies with $H<26.6$. As we assume that more massive galaxies are more likely to be hosts, for each galaxy in the host catalogue, we retain the total stellar mass, in addition to its \ra, \dec, and redshift. 

We build our observed galaxy catalogue to realistically mimic the depth of observations that could be achieved by the next-generation spectroscopic surveys. Hence, we select Flagship galaxies with an apparent magnitude $i < 24.7$, based on the parameters of the future Stage-5 spectroscopic instrument \citep{albrecht_report_2006, besuner_spectroscopic_2025}. 
We also consider redshifts with an uncertainty of $\sigma_z =  \sigma_0(1+z)$ with $\sigma_0 = 0.003$ relative to the true Flagship redshifts. This assumption is conservative with respect to the expected Euclid spectroscopic performance, for which $\sigma_0 < 0.002$ \citep{euclid_calibration}.
The resulting observed galaxy catalogue possesses a line-of-sight density of $4.13\ {\rm galaxies/ arcmin}^2$.

\subsection{Generation of the SGWB energy density}
\label{sec:simulations_omega}
The SGWB is characterised by its energy density $\Omega_\GW$, which represents the fractional contribution of the SGWB to the present-day critical density $\rho_{c,0}$. It can be written as a function of frequency and, in the anisotropic case, angular dependence \citep{allen_detecting_1997, allen_detection_1997}: 
\be \label{eq:omega}\Omega_\GW(f_o, \ve) = \frac{f_o}{\rho_{c,0}}\frac{\dd^3\rho_\GW}{\dd f_o \dd^2 \omega} \ , \ee
where $f_o$ is the observed gravitational wave frequency, $\ve$ the line of sight vector, $\omega$ the infinitesimal solid angle, $\rho_\GW$ the total gravitational wave energy density, and $\rho_{c,0} = 3H_0^2c^2/8\pi G$ the critical density today, with $H_0$ being the Hubble constant.

Equation \ref{eq:omega} can be expressed in terms of an isotropic monopole and its angular fluctuations: 
\be 
\label{eq:omega_iso_aniso_separation} \Omega_{{\rm GW}}(f_o, \ve) = \frac{\overline{\Omega}_{{\rm GW}}(f_o)}{4\pi} + \delta\Omega_{{\rm GW}}(f_o, \ve) \ . 
\ee

Moreover, $\Omega_\GW$ can be decomposed into contributions from different sources. In this study, we consider three classes of \acp{CBC} that dominate the SGWB \citep{capurri_intensity_2021, rosado_gravitational_2011}: binary black holes (BBH), neutron star black holes (NSBH), and binary neutron stars (BNS). We express $\Omega_\GW$ in terms of their merger rate $R^{[i]} \equiv R^{[i]}(z,\vtheta) $, and the intrinsic energy emitted by each merger $\dd^3 E_\GW^{[i]} / \dd f \dd^2 \omega$, both dependent on the binary type, redshift $z$ and intrinsic binary parameters $\vtheta$, such as mass and spin \citep{bellomo_class_gwb_2022, capurri_intensity_2021}:
\be    \label{eq:omega_merger_rate} \Omega_\GW(f_o, \ve) =  \sum_{[i]\in \{{\rm CBC}\}} \frac{f_o}{\rho_{c,0}} \int \dd z \ \dd\vtheta\ \frac{R^{[i]}(z, \ve, \vtheta)}{H(z)(1+z)}\frac{\dd^3 E^{[i]}_\GW}{\dd f_e \dd^2 \omega}(f_o, \vtheta) \ .  
\ee
We consider the energy emission $\dd^3 E_\GW^{[i]} / \dd f_e \dd^2 \omega$ directionally independent. The dependence on the direction of observation is encapsulated in the merger rate. By doing so, we also assume that the energy density is a factorisable function in terms of the frequency and the line of sight direction \citep{allen_detection_1997, romano_detection_2017}. We will study the SGWB at frequency $f = 25$ Hz, where this assumption has been proven suitable \citep{cusin_properties_2019}. 

To simulate the SGWB as a function of observation time $T$, the integrated merger rate over volume and compact binary parameters is equivalent to summing over to the number of events of each type occurring within a time $T$ \citep{bellomo_class_gwb_2022}:
\be  
  \int \dd r\ \dd \vtheta \  4\pi r^2 R^{[i]}(z(r),\vtheta) \rightarrow \frac{1}{T} \sum_{j = 1}^{N^{[i]}} \ , 
\ee
where $r$ is the comoving distance, and the dependence in $\vtheta$ is equivalent to accounting for each merger $j$ individually. The intrinsic energy emitted by each event follows \citep{buonanno_stochastic_2005, marassi_gravitational_2009, phinney_practical_2001}: 
\be 
\label{eq:event_energy_emission} \begin{split}
    & \frac{\dd^3 E_\GW^{[i], j}}{\dd f_e\dd^2 \omega}(f_o, z) =\\& = \frac{4\pi c^3}{8G} \left(\frac{d_L}{1+z}\right)^2 f_o^2\ (|\Tilde{h}_+^{[i], j}(f_o, z)|^2 + |\Tilde{h}_\times^{[i], j}(f_o, z)|^2) \ , \end{split} 
\ee
where $\Tilde{h}_{+/\times}^{[i], j}$ are the Fourier transforms of the gravitational wave strain amplitudes from each merger.
We compute these using the waveform models \texttt{IMRPhenomXP} for BBH and NSBH \citep{pratten_computationally_2021}, and \texttt{IMRPhenomXP\_NRTidal\_v2} for BNS \citep{dietrich_improving_2019}.

Finally, the expression for $\Omega_\GW$ used in our simulations becomes: 
\be \label{eq:omega_amplitude_dependence}
    \begin{split}
    &\Omega_\GW(f_o, \ve ) = \\
    &=\sum_{[i]\in \{{\rm CBC}\}}\sum_{j = 1}^{N^{[i]}}   \frac{c^2}{8G\rho_{c,0}}
    \frac{f_o^3}{T}(|\Tilde{h}_+^{[i],j}(f_o)|^2 + |\Tilde{h}_\times^{[i], j}(f_o)|^2)\delta(\ve - \ve_j) \ , \end{split} 
\ee
where we consider that the amplitude of GWs is independent of the direction of observation, so that $\Omega_\GW(f_o, \ve )$ is a separable function in terms of its frequency and directional dependence. Regarding the latter, it is encapsulated by $\delta(\ve - \ve_j)$, which is a Dirac delta function independent of the frequency. 

To simulate the SGWB using Equation \ref{eq:omega_amplitude_dependence}, we model both the merger rate evolution and the source parameter distributions
using the most recent population inferences from GWTC-4.0  
\citep{abac_gwtc-40_2025, abac_upper_2025}. 
For BBHs, the primary mass distribution is described by the so-called `Broken Power-Law + Two Peaks' model \citep{callister_parameter-free_2024}, while the mass ratio follows a power law distribution \citep{abac_gwtc-40_2025}. For BNS and NSBH systems, the limited number of direct detections motivates the use of simplified models from \cite{abac_upper_2025}. Specifically, we assume a log-uniform distribution between 3 and 50~$M_\odot$ for the black hole mass in NSBH binaries and a uniform distribution between 1 and 2.5 $M_\odot$ for neutron stars in both NSBH and BNS systems. 
The measured effective spin on BBHs is small \citep{abac_gwtc-40_2025}.  At 25 Hz, it can cause an uncertainty of approximately 3\% on the SGWB amplitude \citep{ebersold_uncertainty_2026}. For simplicity, and because the impact on the SGWB is minor, we assume that all binaries are non-spinning.

\begin{figure*}
\includegraphics[width = \linewidth]{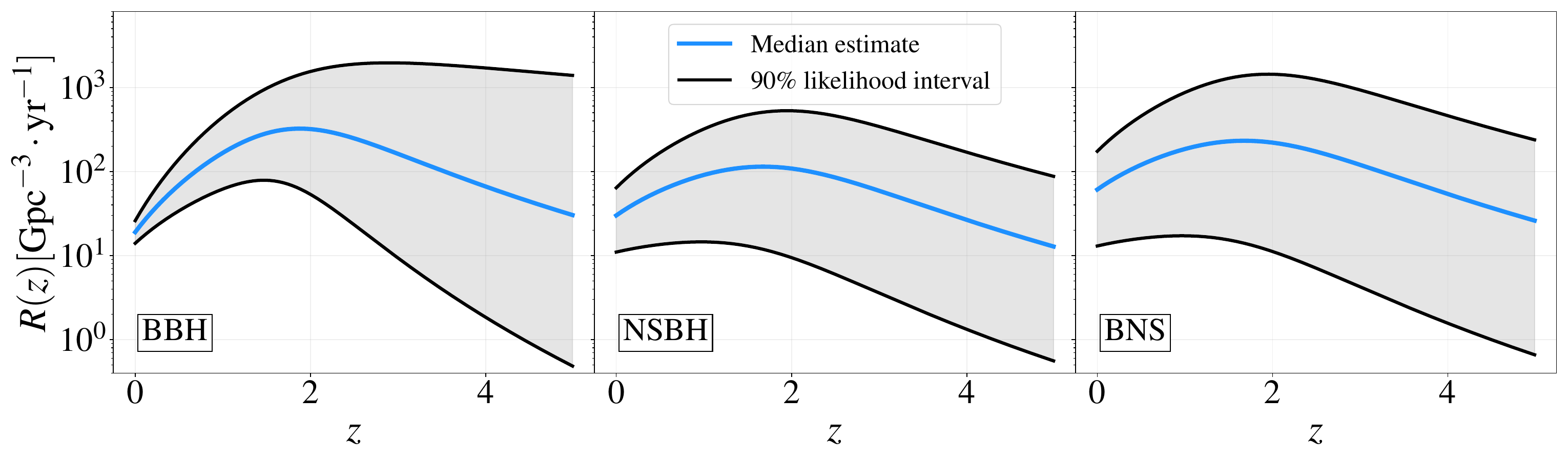}
\caption{Merger rate history estimates for BBH, NSBH and BNS. These estimates are obtained using the broken power law model from Equation~\ref{eq:MD} with parameters given in Table~\ref{tab:MD_param}. The lower and upper estimates are defined from the 90\% likelihood of the O4a LVK run \citep{abac_gwtc-40_2025}.}
\label{fig:merger_rates}
\end{figure*}

CBC events are simulated until $z=3$, as this is the limit of the host galaxy catalogue.
Their merger rate evolution is modelled as a broken power law \citep{callister_shouts_2020}:
\be
\label{eq:MD} R^{[i]}(z) = C(\alpha, \beta, z_{{\rm peak}}) R_0^{[i]}\frac{(1 +z)^\alpha}{1+\left(\frac{1+z}{1+z_{\rm {peak}}}\right)^\beta} \ , 
\ee
where $\alpha$, $\beta - \alpha$ represent the low and high redshift slopes, $z_{{\rm peak}}$ denotes the redshift at which the merger rate peaks, and $C(\alpha, \beta, z_{{\rm peak}}) =1+\left(1+z_{\rm {peak}}\right)^{-\beta}$ is a normalisation constant ensuring $R^{[i]}(0) = R_0^{[i]}$. 
The parameters $z_{\rm {peak}}, \alpha$ and $\beta$ have been constrained by the latest LVK observing run \citep{abac_gwtc-40_2025, abac_upper_2025}. Their median values and 90\% credible intervals are reported in Table \ref{tab:MD_param} and illustrated in Figure \ref{fig:merger_rates}. For BBHs, these parameters are inferred directly from observations, whereas for BNS and NSBH systems only the local rate is observationally constrained, their redshift evolution is modelled following the Madau-Dickinson star formation rate \citep{madau_cosmic_2014} convoluted with a time delay distribution between formation and merger as in  \cite{ebersold_uncertainty_2026} and expressed in terms of Equation \ref{eq:MD}.

\begin{table}
    \centering
    {\renewcommand{\arraystretch}{1.3}
    \begin{tabular}{|l|c|c|c|c|} \hline
        & $R_0^{[i]}\ [{\rm Gpc}^{-3}  {\rm yr}^{-1}]$ & $z_{\rm {peak}}$ & $\alpha$ & $\beta$ \\ \hline
        BBH & $19^{+7}_{-5}$ & $2.0^{+0.2}_{-0.2}$ & $3.2^{+1.0}_{-1.0}$ & $7.6^{+2.0}_{-2.0}$ \\ \hline
        BNS & $61^{+113}_{-48}$   &  \multirow{2}{*}{$2.12^{+0.06}_{-0.21}$} & \multirow{2}{*}{$1.7^{+0.7}_{-1.13}$} & \multirow{2}{*}{$5.94^{+0.32}_{-0.42}$} \\ \cline{1-2}
        NSBH & $30^{+34}_{-19}$  & & &   \\ \hline
    \end{tabular}}
    \caption{Parameters for the source frame merger rate evolution  for each binary type as expressed in Equation \ref{eq:MD}. These yield the same 90\% credible rates as in \citet{abac_upper_2025}.}
    \label{tab:MD_param}
\end{table}

\subsection{Construction of the SGWB sky maps}

To construct the SGWB sky map, we assign the generated compact binaries to galaxies from the host catalogue.
Specifically, we consider each binary within a thin redshift slice $[z, z+\dd z]$ and assign it to a galaxy belonging to the same slice, thereby inheriting the galaxy's right ascension and declination.
The host galaxy is selected according to a stellar mass-weighted probability distribution as more massive galaxies are more likely to host a merger.
This procedure preserves the merger rate evolution (Equation \ref{eq:MD}, Figure \ref{fig:merger_rates}) ensuring a consistent treatment of its uncertainties while using the galaxy population to determine the angular distribution of compact binaries on the sky. 
A direct sky map of individual mergers would implicitly assume that all sources are resolved, and would therefore not correspond to a stochastic background. Instead, we construct a map of the accumulated GW energy density $\Omega_\GW(f_o, \ve)$ from CBC events during some observing time window within each sky pixel.

We generate maps using \textsc{HEALPix}
\citep{zonca_healpy_2019, gorski_healpix_2005} with a map parameter $n_{\rm side} = 32$, which corresponds to an angular scale of $\ell_{\rm max} = 3\times n_{\rm side} -1 = 95$ (circa $1.8^\circ$), approximately the maximum resolution expected for the future Einstein Telescope  \citep{branchesi_science_2023, mentasti_et_2021, mentasti_prospects_2023}.

\begin{figure}
     \centering
     \includegraphics[width=1\linewidth]{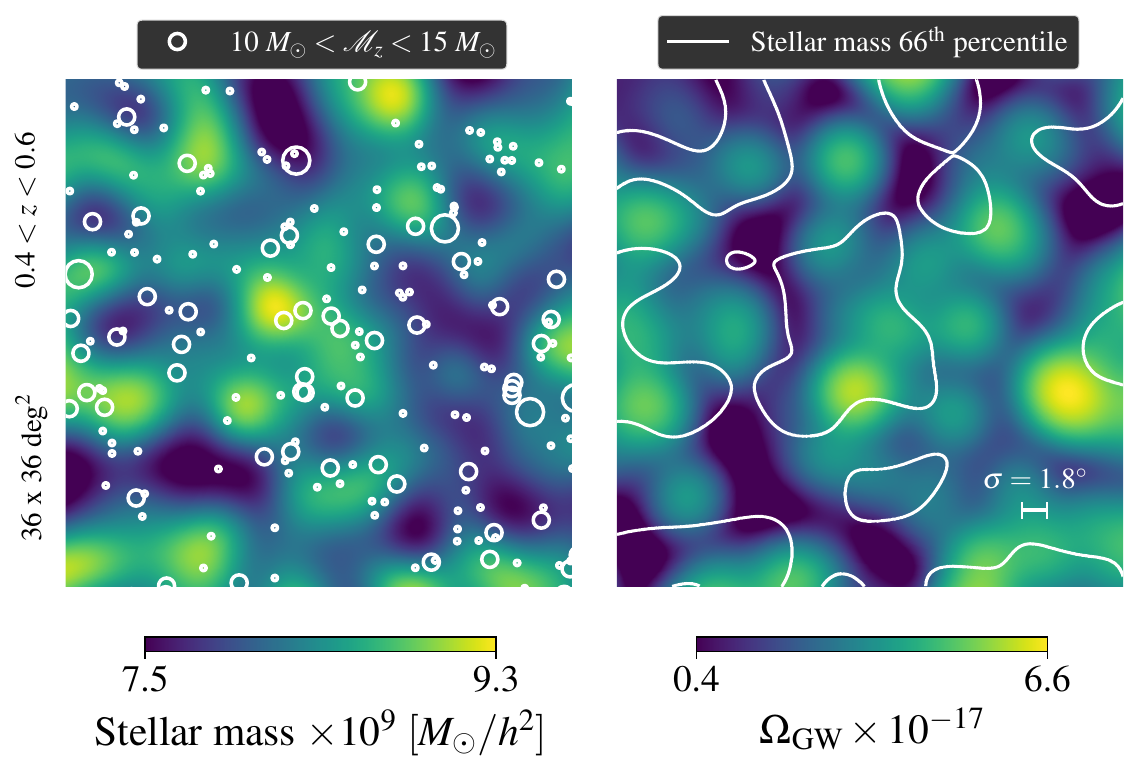}
     \caption{\textit{Left panel:} Distribution of simulated \acp{CBC} overlaid on the total stellar mass of the host galaxy catalogue within the tomographic bin $0.4< z < 0.6$. Each white circle represents an event that occurred in a time of observation $T = 1$ year. The size of the circle corresponds to different chirp mass intervals: $<$10 (small), 10--15 (medium), and $>$15 (large) solar masses. \textit{Right panel:} The gravitational wave energy density, $\Omega_\GW$, produced by the \acp{CBC} displayed on the left panel. The iso-value lines follow the $66^{\rm th}$ percentile of the stellar mass. Both panels cover a field of view of $36\times36\ {\rm deg}^2$, and are smoothed using a Gaussian filter with $\sigma = 1.8^\circ$, corresponding approximately to the smallest angular scale considered in this work.}
     \label{fig:sky_patch}
\end{figure}

Figure \ref{fig:sky_patch} shows the output of our simulation. For illustration purposes, we chose the redshift bin $0.4< z < 0.6$ and a random sky patch of 36 $\times$ 36 deg$^2$. The left panel shows the distribution of \acp{CBC} overlaid on the stellar mass of host galaxies. Each white circle represents an event that occurred in a time of observation $T = 1$ year. Their sizes represent their chirp masses. The right panel displays the distribution of $\Omega_\GW$ released by these \acp{CBC} overlaid with isolines following the $66^{\rm th}$ percentile of the stellar mass distribution. We note that the redshift binning of the \ac{SGWB} is used here only to support the visualization of the generated data. In our subsequent analysis of this simulated data, the redshift binning is applied only to the observed galaxy catalogue, and not to the \ac{SGWB}, as the relevant GW observable is the cumulative energy density sourced by \acp{CBC} across the full cosmic history.

For the analysis, in addition to working with the full redshift range available, we construct the maps of the SGWB energy emission fluctuations as:
\be 
\delta\Omega_\GW(f_o, \ve) = \frac{\Omega_\GW(f_o, \ve) - \overline{\Omega}_\GW(f_o)}{\overline{\Omega}_\GW(f_o)} \ , 
\ee
which assumes that the SGWB monopole $\overline{\Omega}_\GW(f_o)$ is known and detected.
Throughout this work, SGWB anisotropies are simulated at a reference frequency of 25 Hz, corresponding approximately to the peak sensitivity of the LVK detector network \citep{romero-rodriguez_lvk_2024}.

The advantage of generating a pixelised map of $\delta\Omega_\GW(f_o, \ve)$, instead of building it from a spherical harmonics basis, is that more information is retained. 
Generating a map on a spherical harmonics basis would remove extreme values \citep{ainVeryFastStochastic2018}. Moreover, as it will be done in our analysis, one can easily go from a pixelised map to spherical harmonics using the methods implemented in the package \textsc{HEALPix}.  

%% file: estimators.tex
\label{sec:estimators} 
Expressions for the SGWB and galaxy density field angular power spectrum estimators condense the field-level information into 2-point summary statistics. 
To derive such estimators, we first recall that the energy emission of gravitational waves, with its directional dependence given by Equation \ref{eq:omega}, can be decomposed into spherical harmonic multipoles, $Y_{\ell m}(\ve) $: 
\be \label{eq:spherical_dec}\delta\Omega_\GW(\ve , f_o)  = \sum_{\ell,m} \delta\Omega_{\ell m}(f_o) Y_{\ell m}(\ve) \ ,  \ee
where $\delta\Omega_{\ell m}(f_o)$ are their respective coefficients. 

\subsection{Autocorrelation}
\label{sec:estimators_GWB_shot_noise}

The autocorrelation of the SGWB estimator can be expressed as:
\be 
\hat{C}^{{\rm {GW, GW}}}_\ell = \langle\delta\hat{\Omega}_{\ell, m}\delta\hat{\Omega}_{\ell', m'}\rangle |_{\ell = \ell', m= m'} = \frac{1}{2\ell + 1} \sum_{m = -\ell}^\ell \delta\hat{\Omega}_{\ell, m}\delta\hat{\Omega}^*_{\ell, m} \ , 
\ee 
where $\delta\hat{\Omega}_{\ell, m}$ are the coefficients in the spherical decomposition  of the SGWB. These measurements contain the power spectrum plus noise \citep{bellomo_class_gwb_2022, jenkins_shot_2019, tessore_shot_2025}: $\delta\hat{\Omega}_{\ell, m} = \delta\Omega_{\ell, m} + s^{\rm GW}_{\ell, m}$. Hence the autocorrelation is: 
\be  \label{eq:auto_cell_gw}
\begin{split}
    \langle\hat{C}^{{\rm {GW, GW}}}_\ell \rangle = \langle \delta\hat{\Omega}_{\ell, m}\delta\hat{\Omega}_{\ell, m} \rangle =  \langle (\delta\Omega_{\ell, m} + s^{\rm GW}_{\ell,m}) (\delta\Omega_{\ell, m} + s^{\rm GW}_{\ell,m}) \rangle \\
    =   \langle |\delta\Omega_{\ell, m}|^2 \rangle + \langle |s^{\rm GW}_{\ell,m}|^2\rangle = C_\ell^{\rm {GW, GW}} + S^{\rm GW} \ ,
\end{split} 
\ee
where $C_\ell^{\rm {GW, GW}}$ is the autocorrelation signal and $S^{\rm GW}$ is its shot noise.

The shot noise contribution $S^{\rm GW}$ arises from the discrete nature of the \acp{CBC} in both space and time as $\Omega_\GW$ is sourced by individual merger events, each contributing predominantly at the moment of coalescence.
Following the approach of \cite{bellomo_class_gwb_2022}, we derive this shot noise term, often referred to as `popcorn' noise \citep{jenkins_shot_2019}. The only difference is that, in our framework, we compute the probability of hosting a merger within a galaxy rather than within a halo: 
\be 
\label{eq:noise_autocorrelation} \begin{split}
     & S^{\rm GW} =  \ \frac{1}{\overline{\Omega}_\GW^2}\left(\frac{f_o}{\rho_{c,0}}\right)^2 \sum_{[i]\in \{\text{CBC}\}} \int \dd \ve\ \dd z\ \dd \vtheta \\
    & \times \left(\overline{\frac{\dd^3 N_\GW^{[i]}}{\dd z \dd^2 \omega}}\right)^2\ \frac{1 + \beta_T}{\overline{N_\GW^{[i]}}}\left\{\frac{1}{ 4\pi Tc} \left(\frac{1+z}{d_L(z)}\right)^2\frac{\dd^3 E^{[i]}_\GW}{\dd f_e \dd^2 \omega}(f_o, \ve, \vtheta) \right\}^2\ , 
\end{split}
\ee 
where $\overline{N^{[i]}_\GW}\propto T$ and denotes the average number of mergers of type $[i]$ occurring during this observation time $T$, and $\beta_T$ is the mean probability over redshift that a galaxy hosts a merger within the same observation time. 
Equation \ref{eq:noise_autocorrelation} shows that the SGWB shot noise, which will be identified as the main contributor to the cross-correlation covariance, can be interpreted as the number count of merger events weighted by the square of their emitted energy. 

We assume that the SGWB is well described by a Gaussian process. However, the loudest CBC events are outliers that produce strong non-Gaussian fluctuations 
\citep{zhong_importance_2025}. We therefore exclude events with $\Omega_\GW > 10^{-12}$, which amount to less than $1\%$ of the population. This cut suppresses the influence of those outliers while preserving the validity of the Gaussian approximation \citep{sgwb_fluctuations_ginat, sgwb_freq_distribution_ginat}. 
Importantly, our analysis does not rely on a distinction between individually resolved sources and sources contributing to the stochastic background.
Rather, the cut serves as an approximate way of excluding non-Gaussian outliers that would not be considered part of the stationary stochastic signal in a typical cross-correlation SGWB analysis \citep{Renzini-pygwb-2023,abac_upper_2025}. 

\begin{figure}
     \centering
     \includegraphics[width=\linewidth]{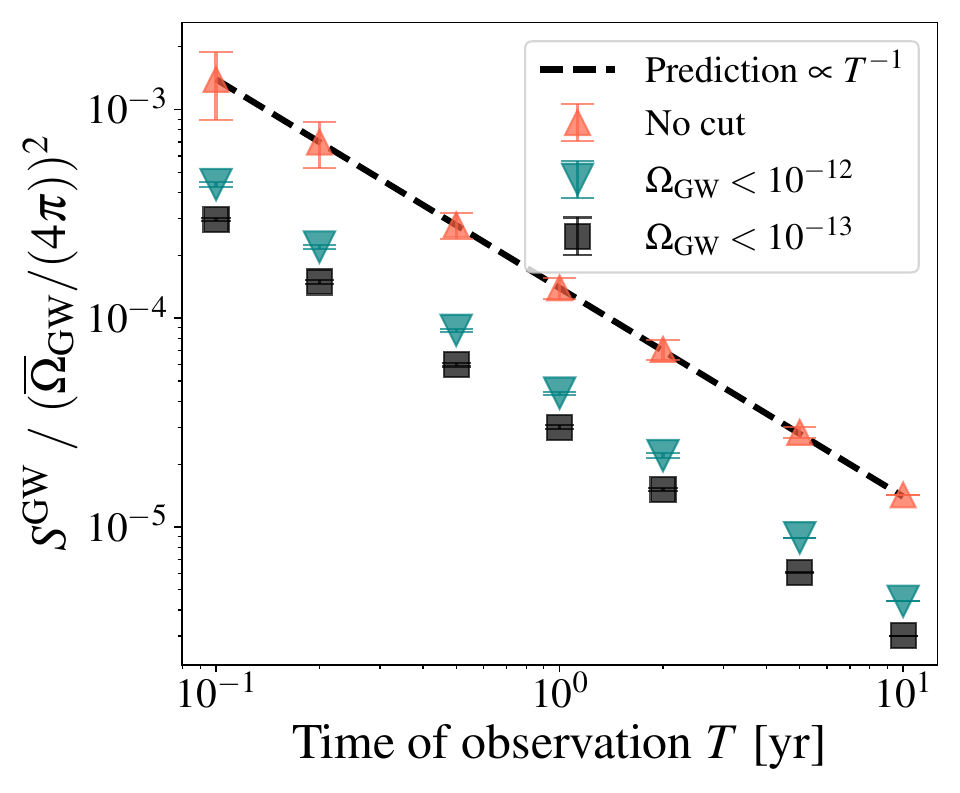}
     \caption{SGWB shot noise, computed from the simulations, with respect to the time of observation is shown for the entire simulated data set as well as after applying the cut-off thresholds  $\Omega_\GW < 10^{-13},\  10^{-12}$. The energy cut-off eliminates the non-Gaussian shot noise contribution of the few loudest events, which are expected to be resolvable and, thus, removable.  Our shot noise computation is also compared with the prediction given by Equation \ref{eq:noise_autocorrelation}, which shows a  $1/T$ dependency.}
     \label{fig:gwb_shot_noise}
\end{figure}

Figure \ref{fig:gwb_shot_noise} compares the numerically estimated shot noise, obtained by autocorrelating randomly distributed compact binary events, with the theoretical prediction of Equation \ref{eq:noise_autocorrelation}, and illustrates its suppression when the loudest events are removed. In addition to the expected $1/T$ behaviour, we see that the energy density threshold suppresses the noise. However, it has diminishing impact for a threshold below $ 10^{-12}$. This justifies our choice of threshold for this paper.

\subsection{Multi-messenger cross-correlation of the SGWB with the galaxy distribution}
Similarly to the autocorrelation, the cross-correlation estimator takes the mean value of the spherically decomposed coefficients of the two tracers, namely the SGWB ($\delta\hat{\Omega}_{\ell, m}$) and the galaxy field ($\hat{\Delta}_{\ell', m'}$):
\be 
\hat{C}^{{\rm {GW, g}}}_\ell = \langle\delta\hat{\Omega}_{\ell, m}\hat{\Delta}_{\ell', m'}\rangle |_{\ell = \ell', m= m'} = \frac{1}{2\ell + 1} \sum_{m = -\ell}^\ell \delta\hat{\Omega}_{\ell, m}\hat{\Delta}^*_{\ell, m} \ . \label{eq:cross_naive} 
\ee

The noise coefficients are not independent, since \acp{CBC} occur inside galaxies. This gives rise to a cross-correlation shot noise term $S^{{\rm {GW, g}}}$:
\be
\langle s^{\rm GW}_{\ell,m}s^{\rm g*}_{\ell,m}\rangle \neq \langle s^{\rm GW}_{\ell,m} \rangle \langle s^{\rm g}_{\ell,m}\rangle
\implies \langle\hat{C}^{{\rm {GW, g}}}_\ell\rangle =   C^{{\rm {GW, g}}}_\ell + S^{{\rm {GW, g}}} \ .
\ee

This shot noise, which originates from the CBC distribution inside galaxies, has been previously identified in  \cite{jenkins_shot_2019,canas-herrera_cross-correlation_2020, alonso_detecting_2020}. We refer to it here as the \textit{host galaxy shot noise} (see Appendix \ref{sec:cross_noise_derivation} for a full derivation): 
\be
    \begin{split}
        &S^{\rm {GW, g}} = \frac{1}{\overline{\Omega}_\GW} \sum_{[i]\in \text{\{CBC\}}} \int \dd z \ \dd \ve \overline{ \frac{\dd^3 N_\GW^{[i]}}{\dd z \dd^2 \omega}} \overline{ \frac{\dd^3 N_g^{{\rm host}}}{\dd z \dd^2 \omega}} \frac{1}{\overline{N_g^{{\rm host}}}} W^{\rm g}(z)W^\GW(z)\ ,
    \end{split}
\ee
where  $\dd^3 N^{[i]}_\GW / \dd z\dd^2 \omega$ and  $\dd^3 N_g^{{\rm host}} / \dd z\dd^2 \omega$ are respectively the number of \acp{CBC} and host galaxies per redshift and per solid angle, while $W^{\rm g}$ and $W^\GW$ are the galaxy and SGWB kernels (see Section \ref{sec:modelling} for a detailed description of the kernels).

\subsection{Tomographic binning of galaxies}

We generate galaxy density maps in tomographic bins based on their redshifts. The optimal binning depends on the galaxy number density and redshift precision of the galaxy survey considered, as one must avoid too narrow bins that would increase the galaxy shot noise while ensuring sufficiently fine binning to capture the redshift evolution of the SGWB.  To keep our forecast as general as possible, we choose a representative tomographic bin width of $\Delta z = 0.2$. 
The 
cross-correlation estimator vector is
\be \label{eq:data_vect} {\bf \hat{C}^{{\rm {GW, g}}}_\ell} = [\hat{C}^{\rm GW, g}_{\ell, i} ]_i \ , \ee
where $\hat{C}^{\rm GW, g}_{\ell, i}$ is the angular power spectrum estimator for the $i^{\rm th}$ tomographic bin. The bold notation is used to indicate a vector containing the estimators for several tomographic bins, at fixed $\ell$.
The covariance matrix of the angular power spectrum in Equation \ref{eq:data_vect}, can be expressed as \citep{friedrich_dark_2021}: \\

\begin{widetext}
\be \label{eq:cov_red_bins}
\begin{split}
     {\bf Cov}({\bf \hat{C}^{{\rm {GW, g}}}_\ell }, {\bf \hat{C}^{{\rm {GW, g}}}_{\ell'} }) = [{\rm Cov}(\hat{C}^{{\rm {GW, g}}}_{\ell, i}, \hat{C}^{{\rm {GW, g}}}_{\ell', j})]_{i, j} =\left[\delta_{\ell\ell'}\frac{\delta_{ij}(C^{\rm GW, GW}_\ell + S^{{\rm GW}})(C^{\rm g, g}_{\ell, i} + S^{\rm g}_{i}) + (C^{\rm GW, g}_{\ell, i} + S^{\rm GW, g})(C^{\rm GW, g}_{\ell, j} + S^{\rm GW, g})}{2\ell +1}\right]_{i, j} \ .
\end{split}
\ee
\end{widetext}

With a tomographic slicing featuring redshift uncertainties much smaller than the bin width, the galaxy cross-correlation between the tomographic bins centred in $z_i$ and $z_j$ and its corresponding shot noise are negligible for $i \neq j$. As we perform a full sky analysis, we also consider the $\ell$ moments to be uncorrelated. 
We bin our signal in $\ell$ bands of width $\Delta\ell =8$, which results in a reduction of the variance by a factor $1/\Delta\ell$ \citep{hu_joint_2004, krause_cosmolike_2017}.

\subsection{Cross-correlation across observation-time bins}
\label{sec:estimators_time_of_obs_binning}
Considering that the background is time-dependent, we can bin it over the time of observation and assume that the bins have noise uncorrelated with each other. By doing so, we can cross-correlate these observation-time bins and construct an unbiased estimator for the SGWB auto-correlation. We take $T$ the total observing time, and $\tau$ a fraction of it such that the number of subdivisions is $N_\tau = T / \tau$. Then we derive the first two moments of the SGWB observation-time cross-correlation. As it is done in \cite{jenkins_estimating_2019}, consider the estimator:
\be 
    \label{eq:time_obs_estimator} \begin{split} \Tilde{C}_\ell^{\rm {GW, GW}} &= \frac{2}{N_\tau(N_\tau - 1)} \sum_{i = 1}^{N_\tau -1} \sum_{j> i}^{N_\tau} ({\bf \hat{C}_\ell^{\rm {GW, GW}}})_{i, j} \\ & = \frac{2}{N_\tau(N_\tau - 1)} \frac{1}{2\ell + 1}\sum_{i = 1}^{N_\tau -1} \sum_{j> i}^{N_\tau}\sum_{m = -\ell}^\ell  \delta\hat{\Omega}_{\ell, m}^i\delta\hat{\Omega}_{\ell, m}^{j*} \ , \end{split} 
\ee
where ${\bf \hat{C}_\ell^{\rm {GW, GW}}}$ is the observation-time cross-correlation matrix of observation bins $i, j$. As it is a symmetric matrix, we are only taking the upper half and removing the diagonal coefficients $i=j$, that would generate an auto-correlation shot noise. Hence, we are summing over $N_\tau(N_\tau - 1)/2$ elements, which explains the normalisation factor in front. 

The maps $i, j$ have been created independently and they should not have correlated noise. Nevertheless, they do trace the same observable. Therefore, the estimator is unbiased:
\be 
    \begin{split} \label{eq:time_obs_estimator_mean}
    \langle \Tilde{C}_\ell^{\rm {GW, GW}} \rangle   = & \frac{2}{N_\tau(N_\tau - 1)} \frac{1}{2\ell + 1} \times \\
    &\sum_{i = 1}^{N_\tau -1}\sum_{j> i}^{N_\tau}\sum_{m = -\ell}^\ell \langle (\delta\Omega_{\ell, m}^i + s^{{\rm GW}, i})(\delta\Omega_{\ell, m}^{j*} + s^{{\rm GW}, j}) \rangle \\
    = &\  C_\ell^{\rm {GW, GW}} \ . 
    \end{split} 
\ee
Notice that we are not taking $i =j$ so that there is no noise arising from the same map being autocorrelated: 
\be \langle s^{{\rm GW}, i} s^{{\rm GW}, j} \rangle = \delta_{i, j} S^{{\rm GW}, \tau} \ . \ee
We added the index $\tau$ to indicate that $S^{{\rm GW}, \tau}$ is the SGWB shot noise over the time of observation $\tau$. From the time dependence of the noise we have that $S^{{\rm GW}, \tau} = S^{{\rm GW}, T}\times N_\tau $. Now we express the variance (as the covariance is diagonal) of the estimator in Equation \ref{eq:time_obs_estimator} (see \cite{jenkins_estimating_2019} for a detailed derivation):

\be 
    \begin{split}\label{eq:var_cross_obs_bins} &{\rm Var}(\Tilde{C}^{{\rm {GW, GW}}}_\ell) \\ &= \frac{2}{2\ell + 1} \left( |C_\ell^{\rm {GW, GW}}|^2 + 2 \frac{C_\ell^{\rm {GW, GW}}S^{{\rm GW}, \tau}}{N_\tau} +2\frac{(S^{{\rm GW}, \tau})^2}{N_\tau(N_\tau-1)} \right) \ \end{split}.
\ee

Note that using this unbiased estimator, we have a slightly higher variance than with the biased estimator obtained from the direct auto-correlation of the same map. For a large number of observation bins, Equation \ref{eq:var_cross_obs_bins} converges to the auto-correlation variance.
Also note that this time binning method is only applicable to the pure SGWB cross-correlations. Using the same method in multi-messenger cross-correlation estimators and their covariance would not affect the noise in that case. 

\subsection{Signal to noise ratio}
\label{sec:estimators_snr}
 
We define the signal-to-noise ratio (SNR) as
\be 
    \label{eq:snr_cross_redshift_bins}({\rm SNR}_{\text{GW, g}})^2 = \sum_{\ell, \ell' = \ell_{\rm min}}^{\ell_{\rm max}} ({\bf \hat{C}^{{\rm {GW, g}}}_\ell })^{\rm T} {\bf Cov}({\bf \hat{C}^{{\rm {GW, g}}}_\ell }, {\bf \hat{C}^{{\rm {GW, g}}}_{\ell'} })^{-1}({\bf \hat{C}^{{\rm {GW, g}}}_{\ell'} }) \ , 
\ee
where the uppercase T means transverse and the sum is over all $\ell$ value centred in each power bin $\Delta\ell$, going from $\ell_{\rm min}$ to $\ell_{\rm max}$.

For the cross-correlation of the SGWB tracer with itself on different observation bins, we have taken the average of all possible cross-correlation of observation bins from the cross-correlation matrix as shown in Equation \ref{eq:time_obs_estimator}. Therefore, we are dealing with a scalar estimator for each $\ell$ value. Considering its variance in Equation \ref{eq:var_cross_obs_bins}, we can express the SNR as: 
\be 
    \label{eq:snr_cross_obs_bins} \begin{split}
    &({\rm SNR}_{\text{GW, GW}})^2 = \\ &= \sum_{\ell, \ell' = \ell_{\rm min}}^{\ell_{\rm max}} \delta_{\ell, \ell'}\ \Tilde{C}_\ell^{ \GW, \GW}\ {\rm Var}(\Tilde{C}^{{\rm {GW, GW}}}_\ell)^{-1}\  \Tilde{C}_{\ell'}^{\GW, \GW}\\ 
    & = \sum_{\ell = \ell_{\rm min}}^{\ell_{\rm max}}\frac{( \Tilde{C}_\ell^{{\rm {GW, GW}}})^2 }{\frac{2}{2\ell + 1} \left\{ |C_\ell^{\rm {GW, GW}}|^2 + 2 \frac{C_\ell^{\rm {GW, GW}}S^{{\rm GW}, \tau}}{N_\tau} +2\frac{(S^{{\rm GW}, \tau})^2}{N_\tau(N_\tau-1)} \right\}} \ .
    \end{split} 
\ee

\subsection{Instrumental noise}
\label{sec:intrumental_noise_methodology}

\begin{figure}
    \centering
    \includegraphics[width=\linewidth]{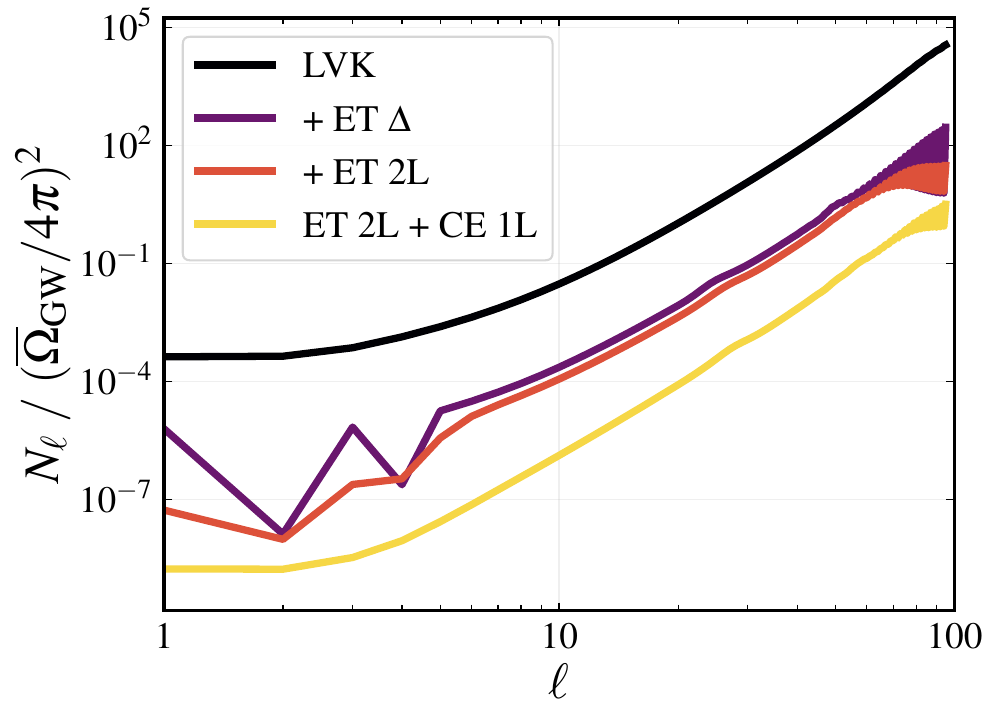}
    \caption{Instrumental noise power spectrum, normalized by the SGWB monopole, computed for the four GW detector network scenarios considered in this paper, assuming 2 years of observation.}
    \label{fig:detector_noise}
\end{figure}

We use the package \textsc{schNell} \citep{alonso_n_ell_2020} to compute the instrumental noise angular power spectrum $N_\ell$ (Figure \ref{fig:detector_noise}) for several possible GW detector network scenarios using  publicly available power spectral density distributions (PSDs): 
\begin{itemize}
\setlength\itemsep{0.5em}
    \item \textbf{LVK} -- The existing LVK detectors, assuming O5 sensitivity 
    \citep{theligoscientificcollaborationAdvancedLIGO2015, AdvancedVirgoSecondgeneration2015a, OverviewKAGRADetector2021}. 
    \item \textbf{LVK + ET $\Delta$} -- One \ac{ET} triangle detector, with an arm length of 10 km, located in the Meuse-Rhine region (latitude $50^\circ\ 43'\ 23''$, longitude $5^\circ\ 55’\ 14''$) \citep{branchesi_science_2023, hild_sensitivity_2011}.
    \item \textbf{LVK + ET 2L} -- Two aligned  L-shaped ET detectors, arm length 15 km, one in the Meuse-Rhine location and one in Sardinia (latitude $40^\circ\ 31’$, longitude $9^\circ\ 25’$) \citep{hild_sensitivity_2011, abacScienceEinsteinTelescope2026, maggiore_science_2020,branchesi_science_2023}.
    \item \textbf{ET 2L + CE 1L} -- The ET 2L above, plus one 40km-long L-shaped CE detector (latitude $46^\circ$, longitude $-125^\circ$)  \citep{Reitze_cosmic_explorer_2019, evans_horizon_2021,guptaCharacterizingGravitationalWave2023}. 
\end{itemize}

To include the detector noise in our study, we consider the simplified case where the instrumental noise is a Gaussian signal independent of the SGWB. It, therefore, only correlates with itself, in the same way as the shot noise does. As a result, Equation \ref{eq:auto_cell_gw} becomes:
\be 
\begin{split}
    \langle\hat{C}^{{\rm {GW, GW}}}_\ell \rangle &= \langle \delta\hat{\Omega}_{\ell, m}\delta\hat{\Omega}_{\ell, m} \rangle \\ 
    &=  \langle (\delta\Omega_{\ell, m} + s^{\rm GW}_{\ell,m} + n_{\ell,m}) (\delta\Omega_{\ell, m} + s^{\rm GW}_{\ell,m} + n_{\ell,m}) \rangle 
    \\ &=   \langle |\delta\Omega_{\ell, m}|^2 \rangle + \langle |s^{\rm GW}_{\ell,m}|^2\rangle + \langle |n^{\rm GW}_{\ell,m}|^2\rangle
    \\ &= C_\ell^{\rm {GW, GW}} + S^{\rm GW} + N_\ell\ .
\end{split} 
\ee

Similarly, for the cross-correlation of the SGWB with galaxies, Equation \ref{eq:cov_red_bins} is modified by: $C^{\rm GW, GW}_\ell + S^{{\rm GW}} \rightarrow C^{\rm GW, GW}_\ell + S^{{\rm GW}} + N_\ell$. 
Finally, for the cross-correlation of the SGWB across observation time bins, we note the dependency on the time of observation: $N_\ell^{\tau} = N_\ell^{T} \times N_\tau$. This implies that we can simply add the instrumental noise and the shot noise: $S^{{\rm GW}, \tau} \rightarrow S^{{\rm GW}, \tau} + N_\ell^{\tau}$.
Equations,  \ref{eq:var_cross_obs_bins}, \ref{eq:snr_cross_redshift_bins}, and \ref{eq:snr_cross_obs_bins} are modified accordingly.

%% file: prediction.tex
\label{sec:modelling}

\subsection{Analytical expressions for the auto- and cross-correlations}
Contributions to the anisotropies can arise from several phenomena \citep{challinorLinearPowerSpectrum2011, cusin_anisotropy_2017}: density, lensing, velocity, and gravitational redshift.  
Here, we only consider the density contribution which dominates over all others. We express the anisotropic term arising from density following the derivation from  \cite{cusin_first_2018}:
\be 
    \label{eq:delta_omega}\delta\Omega_{{\rm GW}}(\ve, f_o) = \frac{1}{4\pi} \int \dd r \partial_r \Bar{\Omega}_{{\rm GW}}(f_o) b_\GW (r) \delta_{m}(r, \ve) \ , 
\ee
where $\delta_{m}(r, \ve)$ is the matter density perturbation at comoving radius $r$ and line of sight $\ve$, and $b_\GW (r) $ is the SGWB bias with respect to such a density.
This bias quantifies the degree to which the SGWB fluctuations follow those of the dark matter distribution \citep{capurri_intensity_2021, capurri_detectability_2022, scelfo_exploring_2020}, whose values are taken from \cite{peron_clustering_2024}. 
From Equation \ref{eq:delta_omega}, we can express the fluctuations in Fourier modes, which can  be decomposed in spherical harmonics:

\be 
    \label{eq:omega_ell} \delta\Omega_{\rm GW,\ell}(k, f_o; \vtheta) = \frac{1}{4\pi}\int \dd r \partial_r \overline{\Omega}_{{\rm GW}}(f_o; \vtheta) (b_\GW (r)\delta_{m,k}(r)j_\ell(kr)) \ . 
\ee
The term  $\partial_r \Bar{\Omega}_\GW$ is  
the kernel of our observable and is defined  as:
\be
    \label{eq:kernel}W(r, f_o) := \frac{1}{4\pi}\partial_r \Bar{\Omega}_\GW(f_o) . 
\ee

For a given pair of maps, of gravitational wave energy emission and galaxy density distribution, the angular power spectra describing the auto- and cross-correlations are:
\be \label{eq:cl_gwb_auto} C_\ell^{{\rm GW, GW}} = \frac{2}{\pi}\int \dd k k^2 |\delta\Omega_{\rm GW,\ell}(k, f_o; \theta)|^2 \ , \ee
\be C_\ell^{\rm {g, g}} = \frac{2}{\pi} \int \dd k k^2 |\Delta_\ell(k)|^2 \ ,\ee
\be \label{eq:cl_gwb_cross} C_\ell^{{\rm GW, g}} = \frac{2}{\pi} \int \dd k k ^2 \delta\Omega_{\rm GW,\ell}^*\Delta_{\ell}\ . \ee

\subsection{Explicit expression for the kernel}
\label{sec:modelling_kernel}
We aim to express analytically the kernel $W(r, f_o)$ defined in Equation \ref{eq:kernel} to obtain an explicit expression for the SGWB angular power spectrum.
From Equations \ref{eq:omega_merger_rate} and \ref{eq:kernel} we can express the kernel as 
\be 
    \begin{split} \label{eq:kernel_merger_rate} & W(r, f_o) = \frac{1}{4\pi} \frac{\partial z}{\partial r} \partial_z \overline{\Omega}_\GW(f_o) 
    \\ &= \frac{1}{4\pi} \frac{f_o}{\rho_{c,0}} \frac{\partial z}{\partial r}  \sum_{[i]\in \{\text{CBC}\}} \int \dd\vtheta \frac{R^{[i]}(z, \vtheta)}{H(z)(1+z)}\frac{\dd E^{[i]}_{\GW}}{\dd f_e}(f_o, z, \vtheta) \ . \end{split} 
\ee
We assume that the merger rate factors into its redshift and intrinsic parameter dependencies: 
\be R^{[i]}(z, \vtheta) = R^{[i]}(z)\Theta^{[i]}(\vtheta), \ee
so that the compact binary parameter distribution is independent of redshift.

The energy emission of a single merger is also redshift dependent, $ \dd E^{[i]}_{GW}/\dd f_e$, as expressed in Equation \ref{eq:event_energy_emission}.
To make this explicit, we isolate the redshift dependence of the Fourier transform of the strain amplitude: $\Tilde{h}_{+/\cross}$. During the inspiral phase, its scaling can be written as \citep{bellomo_class_gwb_2022, capurri_intensity_2021}
\be 
    \Tilde{h}_{+/\cross} \propto \mathcal{M}_z^{5/6} d_L(z)^{-1} f_o^{-7/6} \ ,  
\ee
where $\mathcal{M}_z = (1+z)\mathcal{M}_e$ is the redshifted chirp mass \citep{cutler_gravitational_1994}. To summarise, the redshift dependence of the energy emission per unit frequency becomes
\be 
    \frac{\dd^3 E_\GW}{\dd f_e\dd^2 \omega}(f_o, z)\propto \frac{(1+z)^{5/3}}{(1+z)^2} f_o^{-1/3}\ . 
\ee
With all $z$-dependencies in Equation \ref{eq:kernel_merger_rate} identified, the remaining frequency dependencies follow Equation \ref{eq:spectral_shape} and the constants come directly from the kernel definition in Equation \ref{eq:kernel}.

It is important to note that the decompositions performed in this section are approximations that allow a simple yet precise enough kernel evaluation. This provides a practical alternative to the more detailed formulations by \cite{cusin_anisotropy_2017, jenkins_anisotropies_2018}.

\begin{figure}
    \centering
    \includegraphics[width=\linewidth]{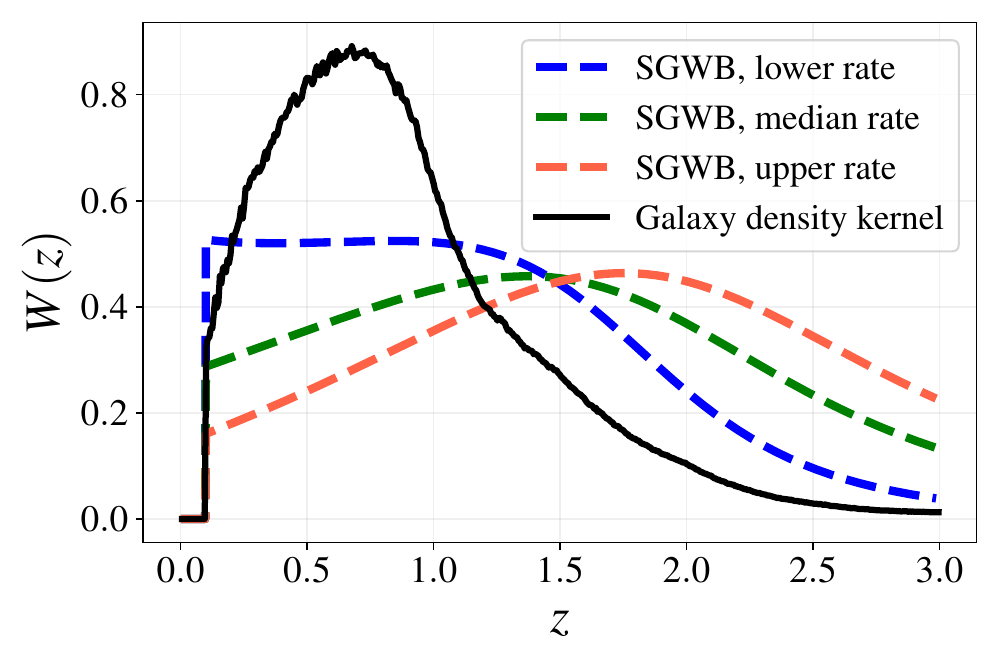}
    \caption{SGWB normalised kernels from the analytical prediction for the three merger rate estimates shown together with the galaxy density kernel. To realistically simulate an observed catalogue, the galaxy density kernel is taken from the Euclid Flagship Catalogue with a magnitude cut $i < 24.7$ and redshift uncertainties $\sigma_z =  \sigma_0(1+z)$, with $\sigma_0 = 0.003$. A lower bound cut is performed for $z < 0.1$ to allow the convergence of the kernel integration.}
    \label{fig:sgwb_galaxy_kernels}
\end{figure}

Figure \ref{fig:sgwb_galaxy_kernels} shows the SGWB kernels obtained analytically for each merger rate estimate considered, and the galaxy density kernel of the observed galaxy catalogue. 
As visible in the figure, the observed galaxy density kernel is negligible as $z\rightarrow3$. Hence, contributions from any CBCs above this redshift would be negligible for a multi-messenger analysis scenario envisioned in this paper, as the galaxy density kernel will suppress the cross-correlation signal.

The redshift cut-off also makes our covariance conservative. As shown in Equation \ref{eq:cov_red_bins}, the covariance includes contributions from the autocorrelation $C_\ell^{\rm GW, GW}$ and the shot noise $S^\GW$. Extending the simulations beyond $z=3$ would add events, lowering $S^\GW$, while mildly affecting  $C_\ell^{\rm GW, GW}$. 
Limiting the CBC population to $z<3$ therefore overestimates the covariance.
This rationale also holds for the cross-correlation across observation-time bins, however with a more significant impact, as the shot noise $S^\GW$ is squared in the variance (Equation \ref{eq:var_cross_obs_bins})

As visible in Figure \ref{fig:sgwb_galaxy_kernels}, the SGWB kernels do not fall-off as $z \rightarrow 0$. This comes from the \acp{CBC} energy-flux dependence $d_L (z) ^{-2}$, which makes the closest sources dominate the SGWB signal. 
Hence, to allow a convergence of the integrals in Equations \ref{eq:noise_autocorrelation} and \ref{eq:cl_gwb_auto}, and therefore a computation of the SGWB auto-correlation and its shot noise, a low redshift cut-off is needed. This is why we consider a redshift lower bound at $z = 0.1$.

The peak in the kernel results in a balance between the evolution of the merger rate, the energy density flux emitted by each merger, and the bias dependence on redshift. This dependence is clearly visible for the lower merger rate estimate when comparing Figures \ref{fig:merger_rates} and \ref{fig:sgwb_galaxy_kernels}. In Figure \ref{fig:merger_rates}, the lower merger rate peaks at $z \sim 1.5$ and starts decreasing at higher redshifts. Thus, the kernel receives a double suppression, from the decrease in the merger rate and the energy-flux dependence $d_L (z) ^{-2}$, which produces for $z > 1.5$ the abrupt fall-off of the kernel (green curve) in Figure \ref{fig:sgwb_galaxy_kernels}.

\subsection{Predicted auto- and cross-correlations}
\begin{figure}
    \includegraphics[width=\columnwidth]{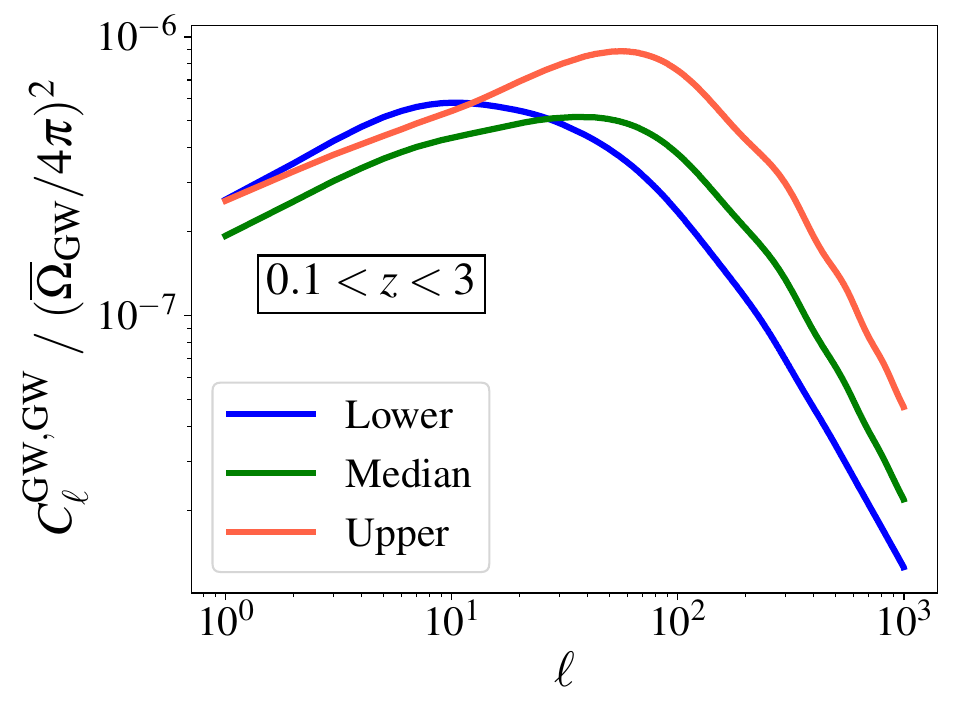}
    \caption{Analytical prediction for the autocorrelations of the SGWB at frequency $f = 25$ Hz, taken over the redshift range $0.1 < z  < 3$ for each merger rate estimate presented in Section \ref{sec:simulations_omega}. The redshift lower bound allows the kernel integration to converge, while the upper bound is set by the range of the host catalogue.}
    \label{fig:gwb_theory_auto}
\end{figure}

The theoretical predictions for the angular power spectrum for the SGWB autocorrelation, corresponding to Equation \ref{eq:cl_gwb_auto},  are displayed in Figure \ref{fig:gwb_theory_auto}. Each curve corresponds to one of the merger rate estimates that we obtained from the LVK O4a constraints (Figure \ref{fig:merger_rates}). For this calculation, we extended the redshift range to the limit supported by the host galaxy catalogue and applied a low-$z$ cut to ensure numerical convergence of the kernel integration. The higher contribution at small redshift of the lower merger rate kernel in Figure \ref{fig:sgwb_galaxy_kernels} can be related to its higher angular power spectrum at small $\ell$'s in Figure \ref{fig:gwb_theory_auto}, while the upper merger rate estimate peaks at higher redshift and is therefore expected to have a higher contribution on larger $\ell$'s. 

\begin{figure}
    \includegraphics[width=\columnwidth]{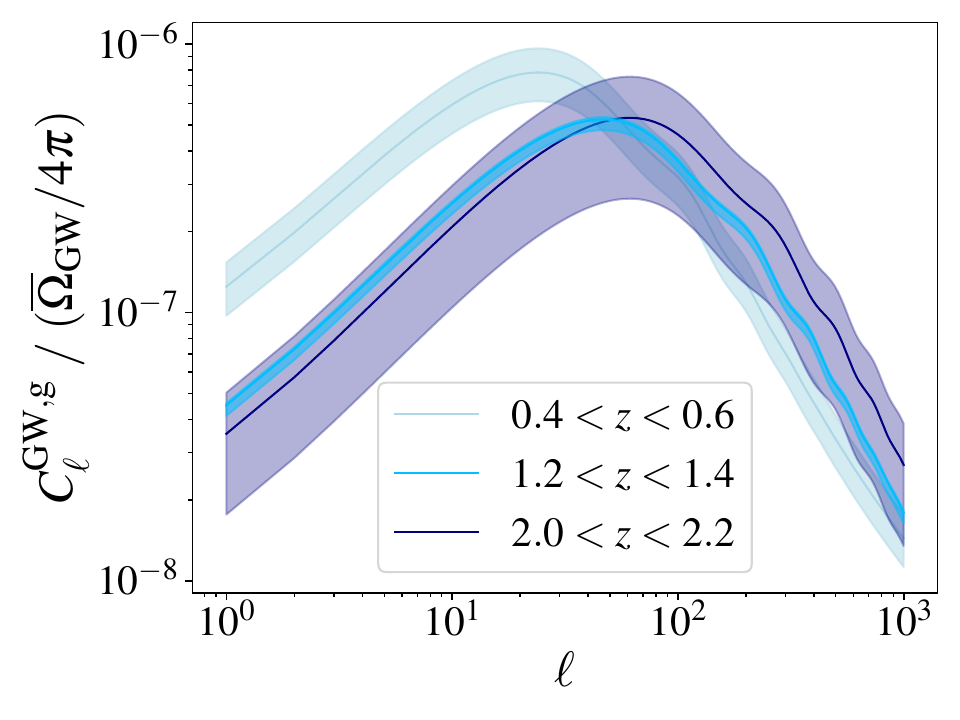}
    \caption{Analytical prediction for cross-correlations of the SGWB, at $f = 25$~Hz, with the galaxy distribution in three representative  bins centred on redshifts $z = 0.5, 1.3$ and $2.1$, with a bin width $\Delta z = 0.2$.
    The shaded area around each curve represents the span covered by the merger rate ranges. The  first redshift bin is the same shown in Figure \ref{fig:sky_patch}.}
    \label{fig:gwb_theory_cross_bins}
\end{figure}

The theoretical predictions for the SGWB cross-correlation with the galaxy distribution, as expressed in Equation \ref{eq:cl_gwb_cross}, are shown in Figure \ref{fig:gwb_theory_cross_bins}. Each curve represents a tomographic bin of the galaxy distribution, while the shaded region indicates the range of possible amplitudes resulting from the different merger rates.
This amplitude spread is very narrow in the central bin and broadens significantly at both low and high redshift. This can be interpreted as a feature of the kernel dependency on the rates. The kernel curves shown in Figure~\ref{fig:sgwb_galaxy_kernels} cross each other at approximately $z\sim 1.5$ and differ significantly at high and low redshifts. 
Moreover, compared to Figure \ref{fig:gwb_theory_auto} where the auto-correlation amplitudes cross each other, in Figure \ref{fig:gwb_theory_cross_bins} the redshift bins are sufficiently narrow so that the cross-correlation amplitudes for the different merger rates do not cross.

\subsection{Covariance estimates}

We compute the covariance matrix associated with the cross-correlation between the SGWB and the galaxy distribution, as defined in Equation \ref{eq:cov_red_bins}, in two ways: first, using the analytical expressions given in this paper to obtain the model prediction, and then using our simulated sky maps. 
In both cases, we consider the diagonal terms only, we bin our signal in $\ell$ bands of width $\Delta\ell = 8$ \citep{hu_joint_2004, krause_cosmolike_2017}.

\begin{figure}
    \centering
    \includegraphics[width=\linewidth]{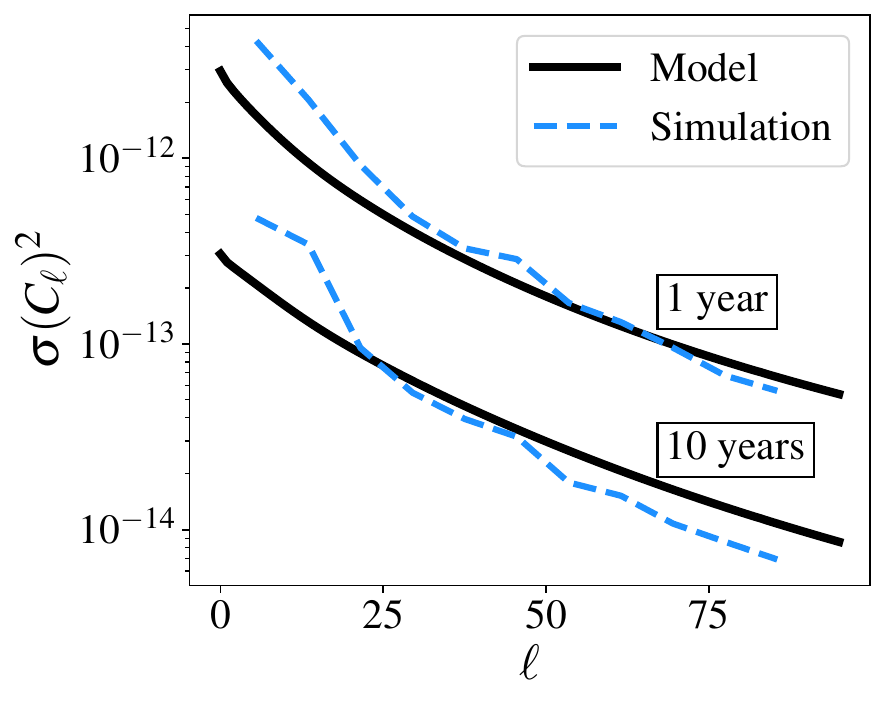}
    \caption{Variance of the angular power spectrum computed, for 1 and 10 years of observation, using the analytical prediction from Equation \ref{eq:cov_red_bins} (solid lines) shown in comparison with estimates from simulations (dashed lines). }
    \label{fig:covariances_obs_times}
\end{figure}

Figure \ref{fig:covariances_obs_times} shows that covariance estimates resulting from the two  computations are in good agreement. As a representative case, we adopted the upper merger rate scenario, for which we show the covariances for observing times of one and ten years, but the agreement holds for all three cases and timescales considered in this paper. The $1/T$ dependence, visible in both approaches, reflects the scaling of the SGWB shot noise.
Mild discrepancies between the two covariance estimates can be seen at the largest angular scales. 
These stem from the construction of the full sky galaxy map from the single Flagship octant, which causes a spurious signal on large scales.

\subsection{Analysis framework}

We use the analytical predictions and simulated data sets to obtain forecasts for the 
detectability 
of both the SGWB cross-correlation in observation-time bins and the multi-messenger cross-correlation in redshift bins. 
Following the convention adopted in the literature \citep{Implication_SGWB_LVK, liImprovingDetectionSensitivity2025, schneiderLowfrequencyGravitationalWaves2001}, we consider the  reference thresholds of  $\rm{SNR}>5$ for \textit{detection} and $3<\rm{SNR}<5$ for \textit{evidence} of a signal.

It is important to not confuse our signal with random noise and show that we have an actual cross-correlation. Therefore, we also compute the cross-correlation of galaxies and compact binaries distributed with random \ra and \dec coordinates. Note that only the angular coordinates are randomized, the galaxy redshift distributions still follow those from the host catalogue and our CBCs follow the merger rates. In each case we study the three merger rate estimates (upper, median, and lower) discussed in Section \ref{sec:simulations_omega}.

We consider the redshift range $0.1 < z < 3$. The lower bound is set so that the kernel integration converges, while the upper bound is set by the limit of the host catalogue. For the multi-messenger analysis, specifically, we also consider bins of width $\Delta z = 0.2$. Representative examples of these bins are shown in Figures \ref{fig:sky_patch} and \ref{fig:gwb_theory_cross_bins}.

We compute the angular power spectra from our analytical prediction in its full form, without the Limber approximation. To do so, we use the FKEM integration method \citep{leonardN5KChallengeNonLimber2023}. 
The use of the full sky simulation has the advantage of ensuring that the $\ell$-modes are uncorrelated resulting in a diagonal covariance matrix in the spherical harmonic basis. 
We consider multipoles in the interval $8 \leq \ell \leq 95$. 
Our limit on small scales, on the one hand, is motivated by the fact that the detector noise increases sharply as $\ell \sim \mathcal{O}(10)$ (Figure~\ref{fig:detector_noise}).
Our limit on large scales, on the other hand, is conservatively set to avoid any spurious signals from the handling of the original simulated sky octant 
(we quantify the impact of this low-$\ell$ cut using the analytical formula).

To compute the cross-correlations from the simulated maps, we use the function \texttt{anafast} from the package \textsc{HEALPix} \citep{zonca_healpy_2019, gorski_healpix_2005}.
To compute the power spectra from analytical predictions, we use \textsc{pyccl} which includes the FKEM integration method 
\citep{chisari_core_2019}.

%% file: results.tex
\label{sec:results}

\subsection{Cross-correlations of the SGWB with the galaxy density}
\label{sec:results_cross}

\begin{figure*}
    \centering
    \includegraphics[width=1.0\linewidth]{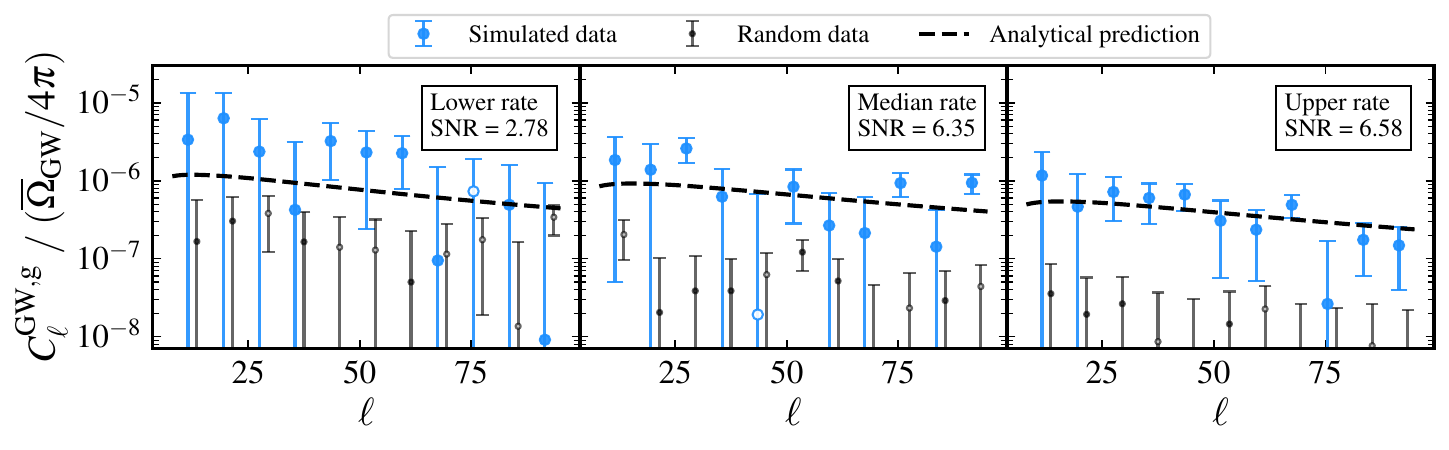}
    \caption{Estimated cross-correlations, in the shot-noise limited regime, between the SGWB and the galaxy distribution in the range $0.1<z<3$ for 2 years of observation (blue) shown in comparison with the expectation from randomly distributed CBCs (black) and the analytical prediction (dashed line). The negative points are given in their absolute value and shown with a white marker face.
    }
    \label{fig:c_ell_upper_random_comparison} 
\end{figure*}

We first consider the shot noise limit case. Figure~\ref{fig:c_ell_upper_random_comparison} shows the multi-messenger cross-correlations in comparison with the expectation from randomly distributed events for 2 years of observation and one wide redshift bin covering the entire range of the simulations. 
Considering the range of angular scales $8\leq\ell \leq 95$, ${\rm SNR} > 5$ is surpassed, by a wide margin,  for the median and upper merger rates. As expected, the lower rate is considerably worse.
Notably, the upper rate is only marginally better than the median rate, due to the fact that, despite having a higher shot noise, the median rate overlaps better with the galaxy distribution than the upper rate (see Figure~\ref{fig:sgwb_galaxy_kernels}).

\begin{figure*}
    \centering
    \includegraphics[width=\linewidth]{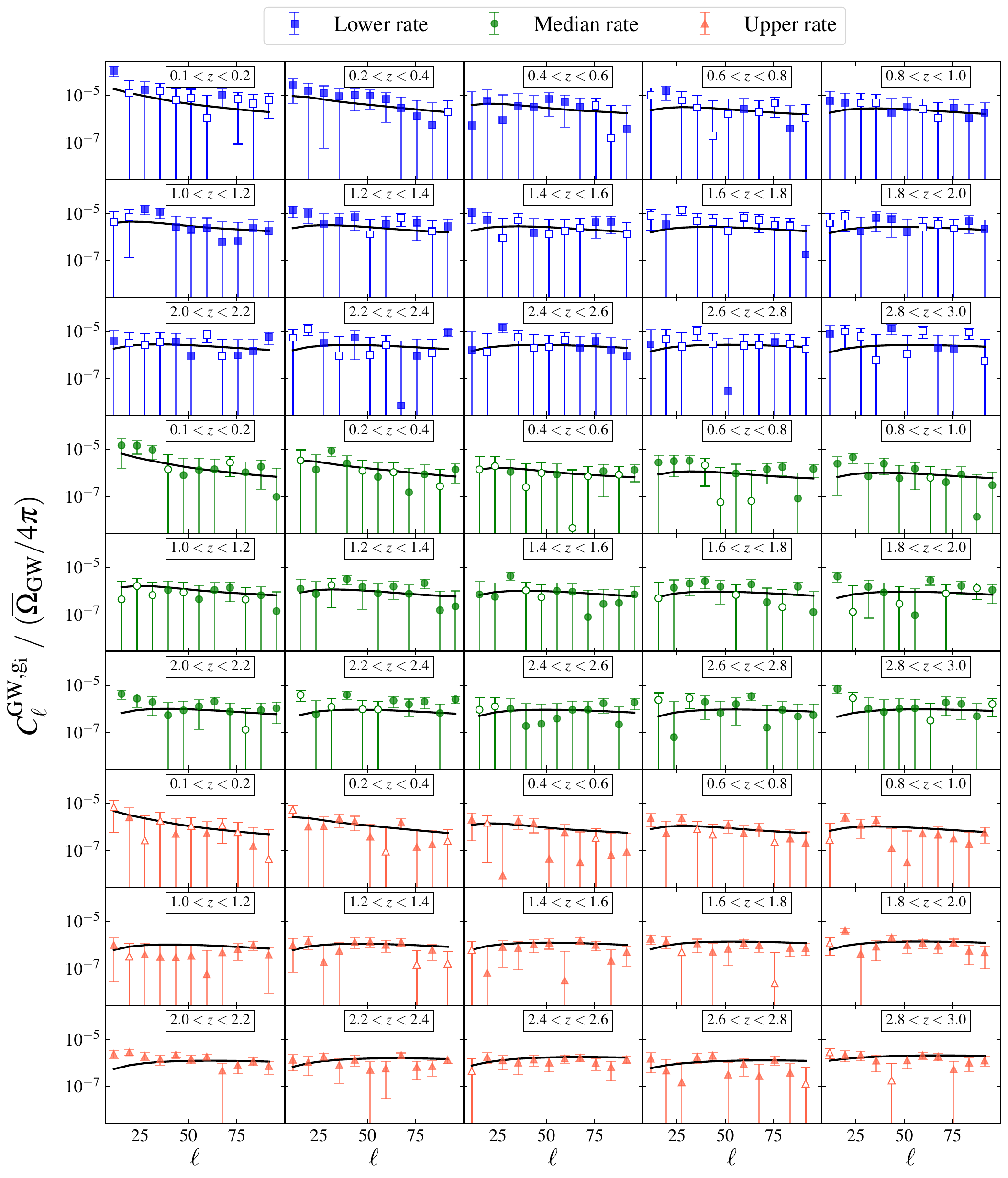}
    \caption{Estimated cross-correlations of the SGWB with the binned galaxy density for 2 years of observation, shown in comparison with the analytical predictions (black lines). The negative points are given in their absolute value and shown with a white marker face. Error bars correspond to shot-noise uncertainties only.}
    \label{fig:cross_correlations_per_bin}
\end{figure*}

\begin{figure}
    \centering
    \includegraphics[width=\linewidth]{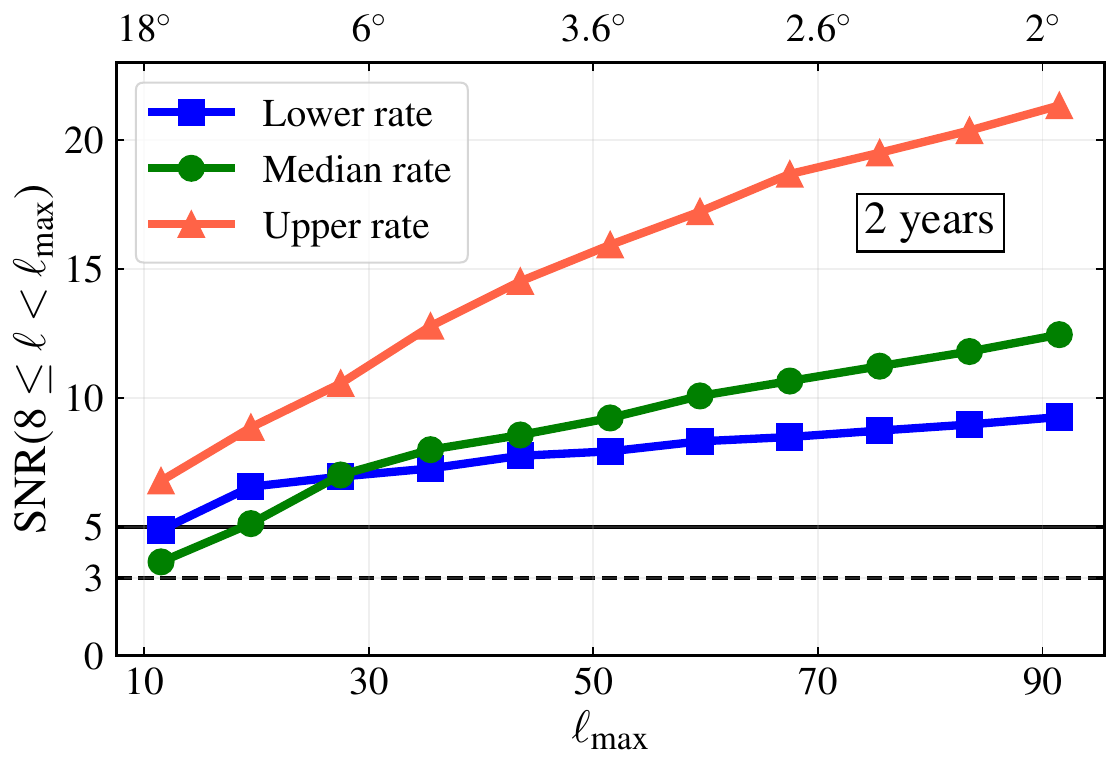}
    \caption{Shot-noise limit SNR calculated using Equation \ref{eq:snr_cross_redshift_bins}, for the three merger rate estimates with respect to the maximal angular scale $\ell_{\rm max}$ considered. The cross-correlation is computed in bins of width $\Delta z = 0.2$, and then combined to get an estimator vector as in Equation \ref{eq:data_vect}, whose covariance is given in Equation \ref{eq:cov_red_bins}. The lines at ${\rm SNR}= 3$ and ${\rm SNR}=5$ are reference thresholds.}
    \label{fig:snr_cmulative_merger_rate}
\end{figure}

\begin{figure}
    \centering
    \includegraphics[width=1\linewidth]{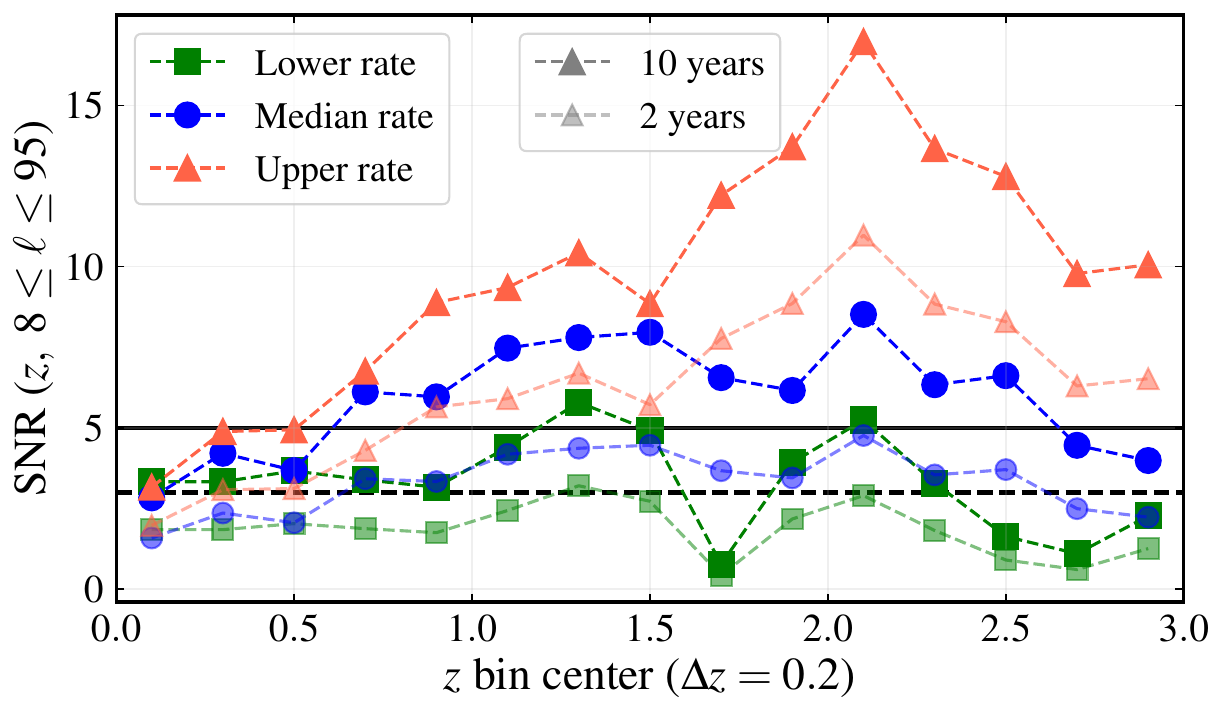}
    \caption{Per-bin shot-noise limit  SNR of the SGWB--galaxy cross-correlation signal. We account only for the signal generated by galaxies within each bin. The SNR is given for each tomographic bin of width $\Delta z = 0.2$ for 2 and 10 years of observation. }
    \label{fig:SNR_cross_correlation_bins}
\end{figure}

We split the galaxies into tomographic bins (Figure \ref{fig:cross_correlations_per_bin}), using an estimator vector as in Equation \ref{eq:data_vect} and a covariance as in Equation \ref{eq:cov_red_bins}. 
Figure \ref{fig:snr_cmulative_merger_rate} shows the resulting shot noise limit SNR  as a function of the maximal $\ell$ value considered, with the horizontal black lines (solid and dashed) indicating the SNR thresholds. We see ${\rm SNR} >3$, for $\ell_{\rm max} = 12 $ and ${\rm SNR} > 5$ for $\ell_{\rm max} = 20 $, for all of the merger rates considered. 
We also see in Figure \ref{fig:snr_cmulative_merger_rate} that the signal for the lower rate is above the median rate in the first two angular bins. Despite its larger covariance, the cross-correlation with the lower merger rate brings a stronger signal. As the lower merger rate kernel has a stronger contribution at low redshift (as visible in the density kernel in Figure \ref{fig:sgwb_galaxy_kernels}), it is expected to have a stronger signal in the low multipoles. This is also depicted by the theoretical angular power spectra in Figure \ref{fig:gwb_theory_auto}. Recalling that the SNR evolves with time as $\sqrt{T}$ (c.f.~Equations 
\ref{eq:noise_autocorrelation}, \ref{eq:cov_red_bins}, and \ref{eq:snr_cross_redshift_bins}), we further  estimate that, for an angular resolution $\ell_{\rm max} = 28$, the signal would be above ${\rm SNR} =5$ in less than 1 year for all merger rate scenarios.  
Figure \ref{fig:SNR_cross_correlation_bins} shows the SNR per redshift bin. Within 2 years, $\rm{SNR}>5$ is achieved for $z > 0.9$ in the upper merger rate scenario and a distinct peak is seen  at $z \simeq 2.0$. This is consistent with the position of the peak of the kernel (Figure~\ref{fig:sgwb_galaxy_kernels}, red curve). After 10 years we have  $\rm{SNR}>5$ for the median rate as well.

\begin{figure}
    \centering
    \includegraphics[width=1\linewidth]{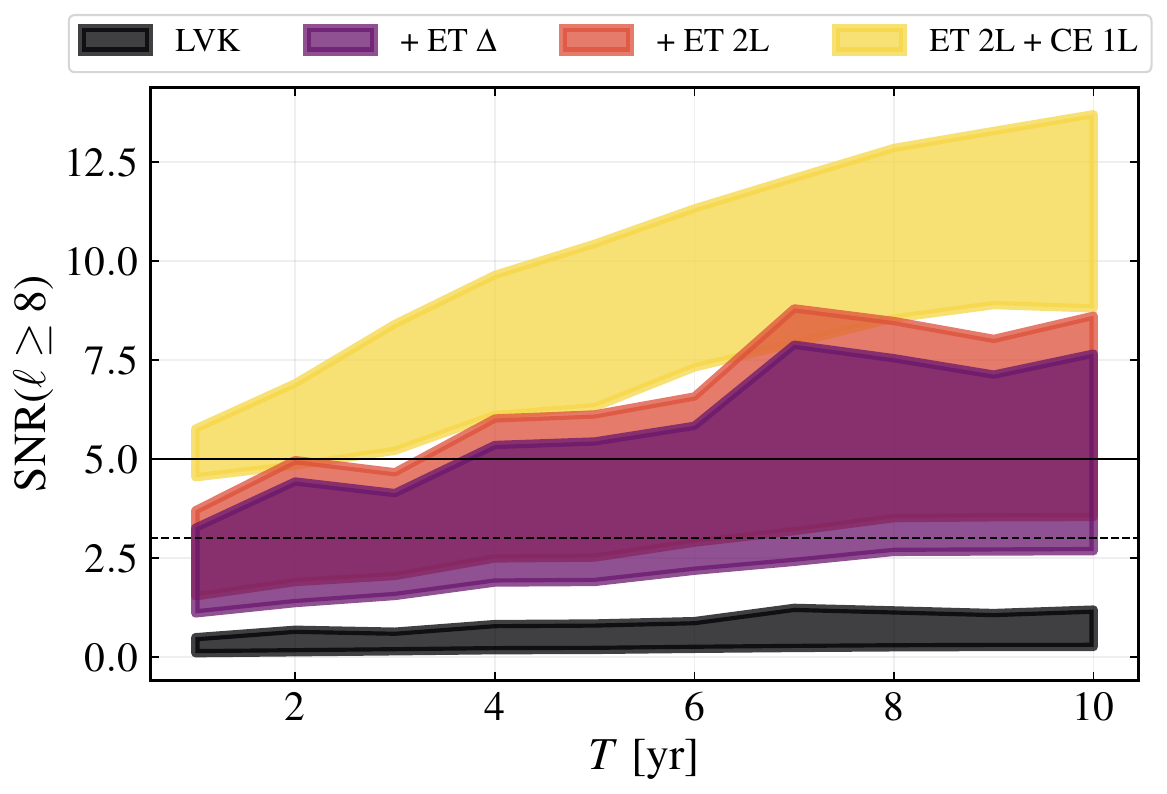}
    \caption{Signal-to-noise ratio with respect to the time of observation computed considering four different networks of detectors. The shaded areas correspond to the SNR range spanned between the three merger rates of reference.}
    \label{fig:snr_detectors}
\end{figure}

We include the instrumental noise in our study by taking the tomographic measurements presented in Figure \ref{fig:cross_correlations_per_bin} and adding to their covariance, in Equation \ref{eq:cov_red_bins}, the noise term $N_\ell$.
The resulting SNR, expected to have a dependence of $\sqrt{T}$, is given in Figure \ref{fig:snr_detectors}, where the coloured bands correspond to the uncertainty arising from the different possible merger rates.
The detectability requirement of ${\rm SNR}>5$ is met after 2 years of observations by a three-detector network formed by ET~2L plus CE~1L regardless of the merger rate assumed.  If merger rates are high,  detection is achieved after just 1 year by such a network.    
For the LVK + ET~$\Delta$ and LVK + ET~2L scenarios, if rates are at the upper bound allowed by the LVK observational constraints, ${\rm SNR}>5$ is achieved after 4 years. If they are at the lower bound, however, the SNR would hover near the evidence threshold of SNR$\simeq$3, even after 10 years. The LVK network at O5 sensitivity does not reach appreciable SNR values.      

Because we previously discarded the scales  $\ell < 8$, our results are conservative. It is important to quantify the impact of our scale cut because, as seen in Figure \ref{fig:detector_noise}, it is at the largest scales that the detectors are the most sensitive. Using our analytical solution,  
we find that the SNR boost from including those large scales does not change our conclusions significantly. For LVK + ET~$\Delta$ and LVK + ET~2L cases, the lower bound of the bands in Figure~\ref{fig:snr_detectors} would remain within $3< {\rm SNR} < 5$.

\subsection{Cross-correlations of the SGWB in observation-time bins}
\label{sec:results_time_bins}

\begin{figure}
    \centering
    \includegraphics[width=\linewidth]{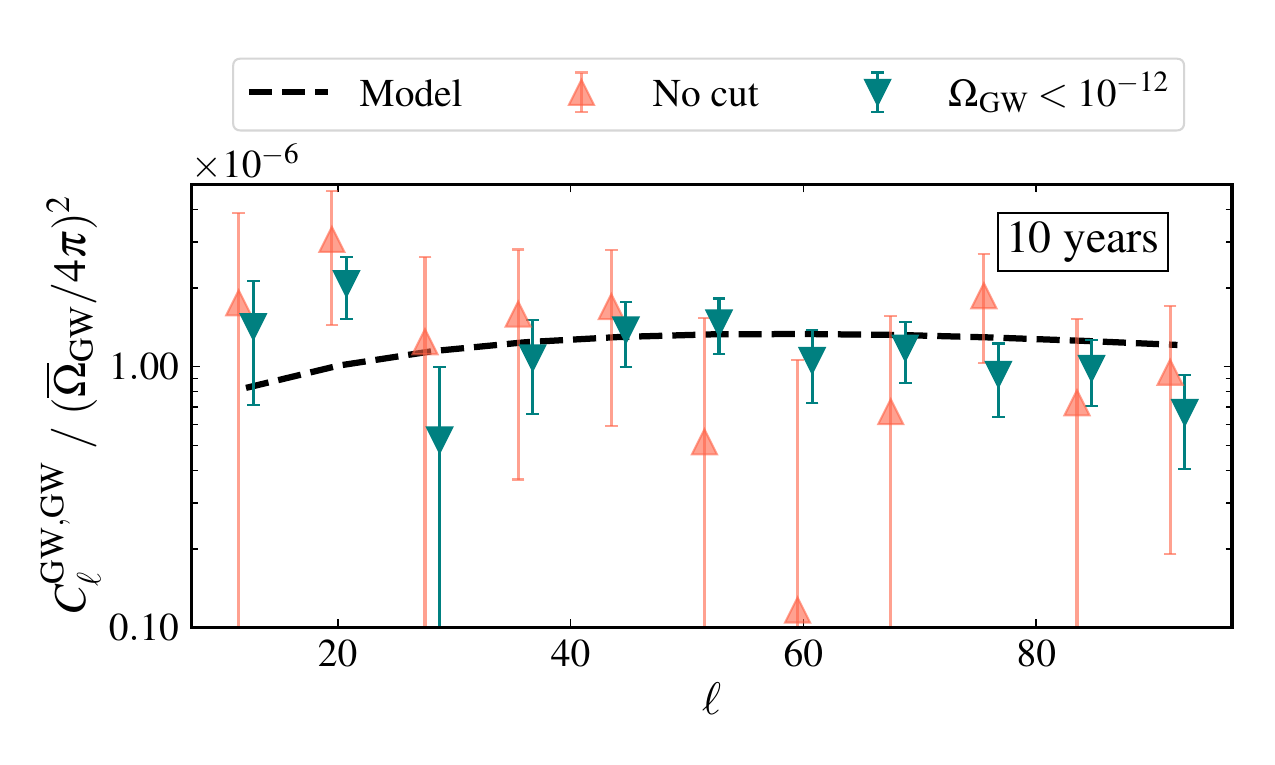}
    \caption{
    Cross-correlation of observation time bins of $\tau = 0.1$ year for a total observation of $T=10$ years. The reconstructed angular power spectrum and their error bars (including shot-noise only) are given for the upper merger rate estimate with and without the energy emission cut-off at $\Omega_\GW < 10^{-12}$. }
    \label{fig:cross_obs_bins_omega_cut}
\end{figure}

The SNR of SGWB cross-correlations across time bins is expected to be low, even in the shot-noise limit regime \citep{alonso_detecting_2020, capurri_detectability_2022}. Therefore, in this section we only analyse the upper merger rate case.
The method employed is based on the discrete temporal nature of events is described in Section \ref{sec:estimators_time_of_obs_binning} (c.f.\ \cite{jenkins_estimating_2019}. 
The cross-correlation covariance is here comparable to the autocorrelation case, which itself is usually larger than the covariance from the multi-messenger cross-correlation.
As shown in Figure \ref{fig:cross_obs_bins_omega_cut},  in which we set the observing time frame to 10 years, we have an unbiased signal estimate of the observation-time cross-correlations.
The dashed black line is the analytical prediction and the error bars are obtained considering only the shot noise. 
While the two results are consistent with each other, we clearly see the improvement of the SNR after the energy cut-off is imposed on the SGWB maps at $\Omega_\GW > 10^{-12}$. 

\begin{figure}
    \centering
    \includegraphics[width=\linewidth]{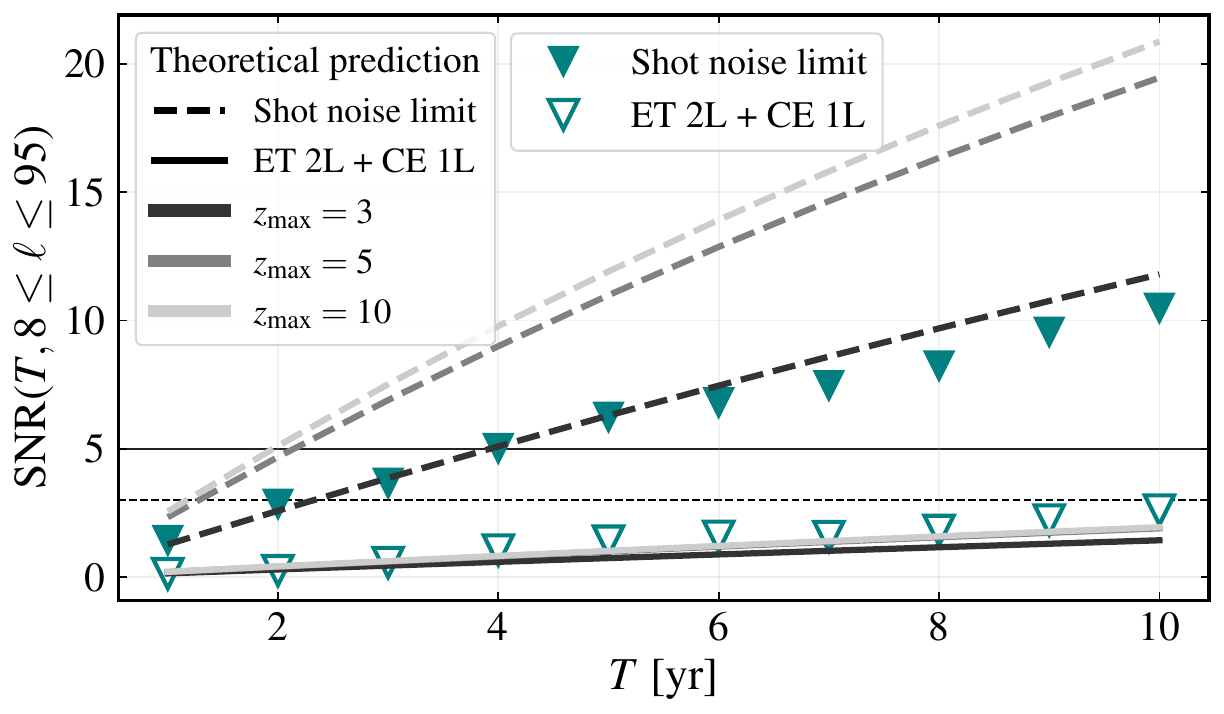}
    \caption{SNR with respect to the total time of observation for the SGWB angular power spectrum reconstructed by cross-correlating year-long observation-time bins. The SNR is calculated using Equation \ref{eq:snr_cross_obs_bins}  and considering only the upper merger rates from Figure~\ref{fig:merger_rates}. For comparison, we show results obtained with and without the instrumental noise for the ET~2L~+~CE~1L detector network.}
    \label{fig:snr_cross_obs_bins_ell_cut_wrt_time_obs}
\end{figure}

The SNR with respect to the time of observation $T$, expected to have a linear dependence, is shown in Figure \ref{fig:snr_cross_obs_bins_ell_cut_wrt_time_obs}, where we also included the instrumental noise for the ET~2L + CE~1L case.
We see that in the shot-noise limit, the SNR exceeds the thresholds ${\rm SNR > 3}$ and ${\rm SNR > 5}$ after 2 and 4 years, respectively. 
Including the expected sensitivity for the ET~2L + CE~1L network considerably dampens the SNR. In contrast with the multi-messenger case, the covariance for the cross-correlation across observation-time contains the instrumental noise term squared (Eq. \ref{eq:var_cross_obs_bins}), which dominates over the angular power spectrum. As a result, in Figure \ref{fig:snr_cross_obs_bins_ell_cut_wrt_time_obs}, the signal lies below the ${\rm SNR} =3$ threshold even after 10 years of observation.
 
Our results here are also expected to be conservative, because we have not considered contributions from CBCs at redshifts  $z>3$ and  excluded $\ell>8$ scales, due to the limits of the input galaxy catalogue. As visible in Figure \ref{fig:sgwb_galaxy_kernels}, the SGWB kernel for the upper merger rate estimate is still significantly above zero at $z=3$, so we expect that CBCs at higher redshifts will have a non-negligible contribution. We quantify how conservative our results are using our theoretical predictions. While CBCs at higher redshifts have a mild impact on the angular power spectrum, they  considerably lower the shot noise.
Therefore, in the shot-noise limit case, the SNR would be considerably enhanced  (grey dashed lines in Figure \ref{fig:snr_cross_obs_bins_ell_cut_wrt_time_obs}) by them. However, when the instrumental noise is included, that enhancement becomes negligible. 
The same conclusion  applies to the scale cut as the instrumental noise erases any gains from including $\ell \leq 8$ scales.

%% file: conclusion.tex
In this paper we  presented forecasts for the SGWB anisotropic signal sourced by a CBC population. 
We considered the multi-messenger cross-correlation of the SGWB with the galaxy distribution and the GW-only cross-correlation across observation-time bins. We simulated SGWB sky maps and developed an analytical model, combining them to obtain empirically grounded forecasts under realistic assumptions including instrumental noise predictions for three GW detector network configurations: ET~2L~+~CE~1L, ET~2L~+~LVK, and ET~$\Delta$~+~LVK. For the galaxies, we assumed a survey with depth $i<24.7$ and redshift uncertainties $\sigma_z = 0.003(1+z)$, well within reach of instruments being developed for upcoming spectroscopic surveys. For the CBC population, we assumed merger rates corresponding to the median and lower/upper 90-percentile range constrained by LVK observations up to O4a.

For the multi-messenger cross-correlation with galaxies, , the ET~2L~+CE~1L network crosses the SNR $=5$ threshold within 2 years of observation (as early as 1 year, if rates are favourable). The other two networks would be limited to $3<$~SNR~$<5$, except for merger rates near the upper limit.  In this case, the SNR threshold would be crossed after approximately 4 years. This result raises the exciting prospect of discovery in the era of third-generation detectors.

The GW-only time bin cross-correlation, in contrast, yields SNR~$<3$ even after 10 years of observation with the most sensitive detector network considered in this investigation.  This result illustrates the challenge of detecting SGWB anisotropies without a multi-messenger tracer.

In addition to those two key findings, further outcomes of this work are the simulation framework and the analytical modelling. First, to generate the simulated SGWB sky maps, we developed and implemented a method to assign CBC events, and their GW energy emission, to an input host galaxy catalogue, selecting the host according to a stellar mass-weighted probability distribution. Our approach was proven accurate in populating a large sample of galaxies covering a wide area of the sky with GW sources and in following their evolution up to the redshift limit of the host catalogue. While we used the Euclid Flagship 2 Galaxy Catalogue as input, this method is general enough to be applied as well to other simulated catalogues. Second, we investigated in detail the properties of estimators for cross-correlations and, more specifically, the behaviour of their inherent noise. We rigorously derived  the SGWB shot noise and confirmed its behaviour with the simulations.
We adapted the SGWB shot-noise formalism to a galaxy hosted population, derived the corresponding cross-correlation shot-noise, and validated both against simulations. This development is directly applicable to the construction of analysis pipelines for observed galaxy catalogues in the future.  
Third, we developed an analytical model of the SGWB kernel based on the redshift-dependent broken power-law used to describe the merger rate. Our model allowed a fast and efficient computation, enabling a complementary perspective to the simulations, and a deeper understanding of the parameter dependencies. The model can be adapted to different merger rates and easily updated for future LVK results.

Our work established to what extent, and in what conditions, multi-messenger cross-correlation are a promising approach for discovery of the astrophysical SGWB anisotropies. We quantified this discovery potential using realistic simulations and detailed theoretical modelling to determine the expected SNR for three possible detector networks in the era of  third-generation GW observatories. In light of the ongoing plans for the Einstein Telescope and Cosmic Explorer, and for future galaxy surveys, this bodes well for the exploration of the SGWB for astrophysics and cosmology investigations.

%% file: appendix.tex
\newpage

\section{Derivation of the host galaxy shot noise}
\label{sec:cross_noise_derivation}

In an analogous way to \cite{bellomo_class_gwb_2022}, we derive an expression for the \textit{host galaxy shot noise} with respect to our compact binary population parameters by integrating over solid angles the expectation value of the relative anisotropic component $\delta \Omega_\GW$ multiplied by the galaxy density:
\begin{widetext}
\be \label{eq:covariance_crosscorrelation} \begin{split}
     C_\ell^{\rm {GW, g}} + S^{\rm {GW, g}} & = \iint \dd \ve \dd \ve' \ \langle \delta\Omega_\GW(\ve) \Delta_{\rm g}(\ve') \rangle \\
    & = \frac{1}{\overline{\Omega}_\GW}\left(\frac{f_o}{\rho_{c, 0}}\right) \sum_{[i]\in \{\text{CBC}\}} \iint \dd \ve \dd \ve' \iint \dd z \dd z' \left\langle\frac{\dd^3 N_\GW^{[i]}}{\dd z \dd^2 \omega} \frac{\dd^3 N_{\rm g}^{{\rm obs}}}{\dd z' \dd^2 \omega'} \right\rangle  W^g(z') \cdot \left\{\frac{1}{ 4\pi Tc} \left(\frac{1+z}{d_L(z)}\right)^2\frac{\dd^3 E^{[i]}_\GW}{\dd f_e \dd^2 \omega}(f_o,z) \right\} \ ,
\end{split}
\ee 
\end{widetext}
\noindent where $\dd^3 N_{g}^{{\rm obs}}/\dd z \dd^2 \omega$ is the density of galaxies in the observed galaxy catalogue, and $W^g(z')$ the galaxy window function used for the cross-correlation. 
The number of events per redshift and solid angle $\dd^3 N^{[i]}_\GW / \dd z\dd^2 \omega$ can be expressed in terms of the average number of events plus a relative variation $\delta_\GW^{[i]}(z, \ve)$ with zero mean value: 
\be \frac{\dd^3 N_\GW^{[i]}}{\dd z \dd^2 \omega} = \overline{\frac{\dd^3 N_\GW^{[i]}}{\dd z \dd^2 \omega}} (1 + \delta_\GW^{[i]}(z, \ve) )\ . \ee
A similar decomposition can be done for the galaxy density. 
 
The mean value of the two densities can be decomposed 
as 
\citep{feldman_power-spectrum_1994}:
\be \begin{split}
    \label{eq:Feldman_decomposition} & \langle (1+\delta_\GW^{[i]}({\bf r}))(1+\delta_g({\bf r'})) \rangle  \\&= 1 + \xi_{\rm GW, g}({\bf r-r'})  + \langle \delta_\GW^{[i]}({\bf r}) \delta_{\rm g}({\bf r'}) \rangle \delta^D({\bf r-r'})\  ,
    \end{split}
    \ee
where $\delta^D({\bf r-r'})$ is a Dirac delta function.
The first term gives the monopole, the correlation function $\xi_{\rm GW, g}({\bf r-r'})$ gives the angular power spectrum, and the third one is the contribution to the shot noise.
This term here does not vanish as both maps are built from the same original host galaxy catalogue:
\be 
    \label{eq:delta_g-delta_gw} \langle \delta_\GW^{[i]} \delta_g \rangle = \left\langle\left( \frac{N_\GW^{[i]} - \overline{N_\GW^{[i]}}}{\overline{N_\GW^{[i]}}} \right) \left(\frac{N_g^{{\rm obs}} - \overline{N_g^{{\rm obs}}}}{\overline{N_g^{{\rm obs}}}}\right)\right  \rangle = \frac{\langle N_\GW^{[i]} N_g^{{\rm obs}} \rangle}{\overline{N_\GW^{[i]}}\overline{N_g^{{\rm obs}}}} - 1 \ .
\ee

Spatially, events are hosted in galaxies whose number $N_g^{{\rm host}}$ follows a Poisson distribution. Temporally, events also follow a Poisson distribution described by the number of events of type $[i]$ per galaxy in a given time of observation $N_{\rm \GW/g}^{[i]}$, whose mean corresponds to the probability that a galaxy hosts a merger of type $[i]$: $\overline{N_{\rm \GW/g}^{[i]}} (z) =  \beta_T^{[i]}(z)$. The total number of events can be written as: 
\be \label{eq:n_gw-n_g} N_\GW^{[i]} = \sum_{k=1}^{N_g^{{\rm host}}} N_{\rm \GW/g}^{[i], k} \ . \ee
Considering that these parameters follow a Poisson distribution, we can infer from the total distribution of events, the two first moments:
\be \label{eq:n_gw-n_g-avg} \overline{N_\GW^{[i]}}  = \beta_T^{[i]} \overline{ N_{\rm g}^{{\rm host}}}\ , \ee
\be 
    \begin{split}
    \langle (N_\GW^{[i]})^2 \rangle &  = \text{Var}(N_\GW^{[i]}) +  \overline{N_\GW^{[i]}} ^2 \\
    & = \overline{N_{\rm g}^{{\rm host}}} \text{Var}(N_{\rm \GW/g}^{[i]})  +  \overline{N_{\rm \GW/g}^{[i]}}^2 \text{Var}(N_{\rm g}^{{\rm host}}) + \overline{N_\GW^{[i]}}^2 \\
    & = \overline{N_{\rm g}^{{\rm host}}} (\beta_T^{[i]} + (\beta_T^{[i]}) ^2)  + \overline{N_\GW^{[i]}}^2  \ , 
    \end{split} 
\ee
where we have used that $ \text{Var}(N_{\GW/g}^{[i]}) =  \overline{N_{\GW/g}^{[i]}} = \beta_T^{[i]} $ and $\text{Var}(N_g^{{\rm host}}) =  \overline{N_g^{{\rm host}}}$, using the properties of a Poisson distribution.

As an approximation, we can assume the number of mergers to be the number of host galaxies multiplied by the the probability of a galaxy hosting a merger: 
\be N_\GW^{[i]} \approx \beta_T^{[i],{\rm eff}} N_g^{{\rm host}} \ . \ee

Note that by taking $\beta_T^{[i],{\rm eff}} $, we are approximating the parameter $\beta_T^{[i],{\rm eff}} $ to a constant over redshift. 
This assumption is reasonable as long as the number of host galaxies is much larger than the number of mergers: $\beta_T^{[i],{\rm eff}}  \ll 1$. 

The number of observed galaxies is also a subsample of the larger host galaxy catalogue, where only a fraction $\alpha$ of galaxies are selected:
\be 
\label{eq:n_g_cat-n_g_host} N_g^{{\rm obs}} = \alpha N_g^{{\rm host}}, \ \alpha \in (0, 1)\ . 
\ee
Hence we approximate Equation \ref{eq:delta_g-delta_gw} as:
\be  
    \label{eq:delta_gw_delta_g-host} \langle \delta_\GW^{[i]} \delta_g \rangle = \frac{\beta_T^{[i],{\rm eff}} \alpha \langle (N_g^{{\rm host}})^2 \rangle}{\beta_T^{[i],{\rm eff}} \alpha\overline{N_g^{{\rm host}}}^2} - 1 . 
\ee

The number of host galaxies follows a Poisson distribution: $\langle (N_{\rm g}^{{\rm host}})^2 \rangle = \text{Var}(N_{\rm g}^{{\rm host}}) + \overline{N_{\rm g}^{{\rm host}}} ^2 = \overline{N_{\rm g}^{{\rm host}}} + \overline{N_{\rm g}^{{\rm host}}}^2 $. We, therefore, simplify Equation \ref{eq:delta_gw_delta_g-host}, 
\be 
    \langle \delta_\GW^{[i]} \delta_{\rm g} \rangle  = \frac{1}{\overline{N_{\rm g}^{{\rm host}}}} \ ,
\ee
and plug it back into the shot noise term of Equation \ref{eq:covariance_crosscorrelation} to obtain:
\be \label{eq:cross_shot_noise}
    \begin{split}
        &S^{\rm {GW, g}} = \frac{1}{\overline{\Omega}_\GW} \sum_{[i]\in \text{\{CBC\}}} \int \dd z \ \dd \ve \overline{ \frac{\dd^3 N_\GW^{[i]}}{\dd z \dd^2 \omega}} \overline{ \frac{\dd^3 N_g^{{\rm host}}}{\dd z \dd^2 \omega}} \frac{1}{\overline{N_g^{{\rm host}}}} W^{\rm g}(z)W^\GW(z)\ ,
    \end{split}
\ee
where we have identified the last factor, arising from the gravitational wave emission, with the SGWB kernel, $W^{\rm GW}(z)$.